\documentclass[10pt,twocolumn,letterpaper]{article}
\PassOptionsToPackage{table,dvipsnames}{xcolor}
\usepackage[pagenumbers]{cvpr} 

\usepackage[ruled,vlined]{algorithm2e}

\def \customparskip {.3em}
\renewcommand{\paragraph}[1]{\vspace{\customparskip}\noindent\textbf{#1}}

\definecolor{turquoise}{cmyk}{0.65,0,0.1,0.3}
\definecolor{purple}{rgb}{0.65,0,0.65}
\definecolor{dark_green}{rgb}{0, 0.5, 0}
\definecolor{orange}{rgb}{0.8, 0.6, 0.2}
\definecolor{red}{rgb}{0.8, 0.2, 0.2}
\definecolor{darkred}{rgb}{0.6, 0.1, 0.05}
\definecolor{blueish}{rgb}{0.0, 0.3, .6}
\definecolor{light_gray}{rgb}{0.7, 0.7, .7}
\definecolor{pink}{rgb}{0.9, 0, 0.6}
\definecolor{greyblue}{rgb}{0.25, 0.25, 1}
\definecolor{teal}{rgb}{0.0, 0.4, 0.4}

\newcommand{\sy}[1]{{\color{dark_green}#1}}

\newcommand{\sample}{{\Phi}}

\renewcommand{\paragraph}[1]{\smallskip\noindent\textbf{#1}}

\definecolor{cvprblue}{rgb}{0.21,0.49,0.74}
\usepackage[pagebackref,breaklinks,colorlinks,allcolors=cvprblue]{hyperref}

\def\paperID{00004} 
\def\confName{CVPR}
\def\confYear{2025}

\title{\emph{Voost}: A Unified and Scalable Diffusion Transformer for Bidirectional \\ Virtual Try-On and Try-Off}

\author{
Seungyong Lee\textsuperscript{1}\thanks{Equal contribution} \hspace{1.5cm}
Jeong-gi Kwak\textsuperscript{1,2}\footnotemark[1]\hspace{1.5mm}\thanks{Corresponding author} \vspace{0.1cm} \\
 \hspace{1.2cm} \textsuperscript{1} NXN Labs \hspace{0.9cm}
  \textsuperscript{2} University of British Columbia
 \\ 
{\small {\url{https://nxnai.github.io/Voost}}}
}

\makeatletter
\apptocmd\@maketitle{{\teaserfigure{}\par}}{}{}
\makeatother

\def\myshift#1{\raisebox{0.5ex}}
\newcommand{\teasernobox}{
\begin{subfigure}[b]{\linewidth}
    \centering
 	\includegraphics[width=\linewidth]{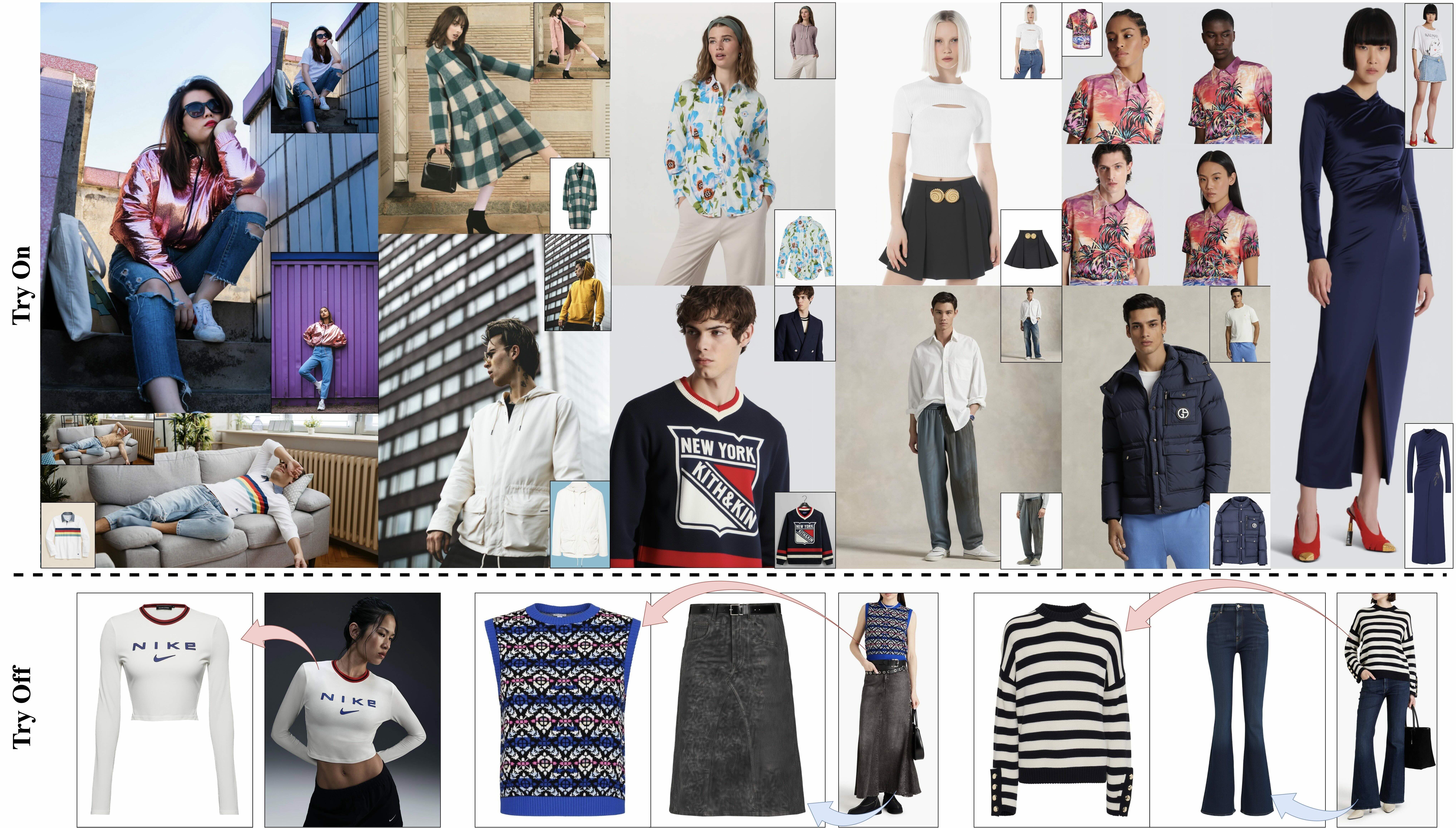}
\end{subfigure}
}

\newcommand{\teaserfigure}{
\vspace{-4mm}
\captionsetup{type=figure}

\teasernobox

\setcounter{figure}{0} 
\captionsetup{type=figure}
\captionof{figure}{
{\bf Teaser -- } 
We present a single diffusion transformer that jointly addresses virtual try-on and try-off, producing high-quality results that remain robust across diverse poses, garment types, backgrounds, lighting conditions, and image compositions.
}
\vspace{4mm}
\label{fig:teaser}
}
\vspace{-1em}

\begin{document}
\maketitle
\begin{abstract}

Virtual try-on aims to synthesize a realistic image of a person wearing a target garment, but accurately modeling garment–body correspondence remains a persistent challenge, especially under pose and appearance variation.
In this paper, we propose \textbf{\emph{Voost}}—a unified and scalable framework that jointly learns \underline{\textbf{v}}irtual try-\underline{\textbf{o}}n and try-\underline{\textbf{o}}ff with a \underline{\textbf{s}}ingle diffusion \underline{\textbf{t}}ransformer.
By modeling both tasks jointly, \emph{Voost} enables each garment-person pair to supervise both directions and supports flexible conditioning over generation direction and garment category—enhancing garment–body relational reasoning without task-specific networks, auxiliary losses, or additional labels.
In addition, we introduce two inference-time techniques: attention temperature scaling for robustness to resolution or mask variation, and self-corrective sampling that leverages bidirectional consistency between tasks.
Extensive experiments demonstrate that \emph{Voost} achieves state-of-the-art results on both try-on and try-off benchmarks, consistently outperforming strong baselines in alignment accuracy, visual fidelity, and generalization.

\end{abstract}

\section{Introduction}
\label{sec:intro}

\begin{figure*}[t]
    \centering
    \includegraphics[width=\linewidth]{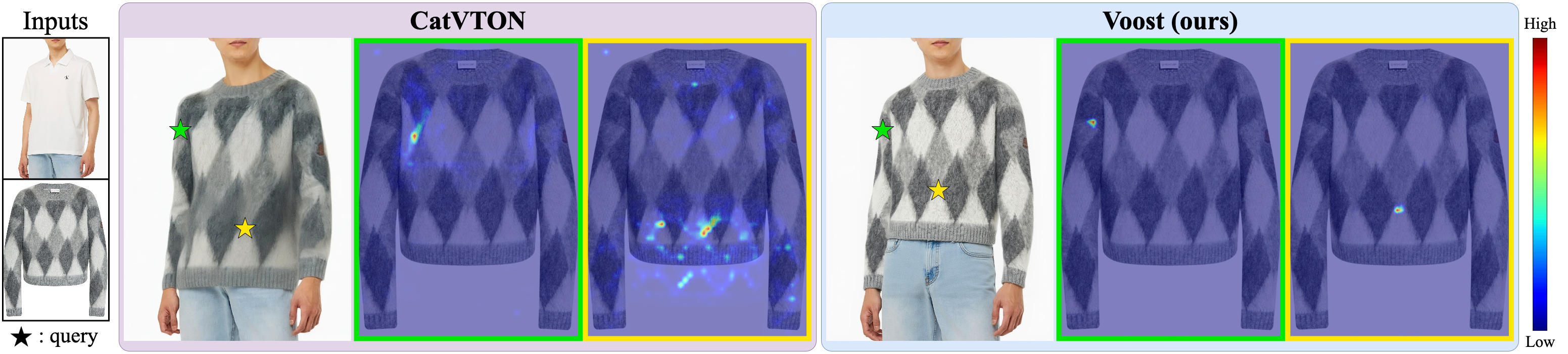}
    \caption{
    \textbf{Attention map comparison—} CatVTON~\cite{chong2024catvtonconcatenationneedvirtual} shows dispersed attention unrelated to the query point, indicating weak spatial grounding. In contrast, our model produces sharply localized maps that align well with the corresponding garment regions, demonstrating stronger relational understanding.
    }
    \label{fig:intro}
\end{figure*}


Virtual try-on (VTON) is an emerging generative task that enables realistic garment transfer onto a person image, offering a new paradigm for fashion in e-commerce and AR/VR.
Despite its promise, VTON remains challenging due to the need for precise alignment, detail preservation, and robustness to occlusion and pose variation.

Early approaches~\cite{Han_2018_CVPR,Yang_2020_CVPR_gan,ge2021parser_gan,choi2021viton} have adopted warping-based strategies, often combined with generative modeling, but have struggled to maintain garment fidelity and structural consistency.

Effectively adapting T2I diffusion models~\cite{rombach2022high,esser2024scalingrectifiedflowtransformers} to virtual try-on requires strong image-based conditioning and accurate modeling of garment--person correspondence. To this end, various efforts have been made to improve spatial alignment and relational understanding. These include incorporating auxiliary objectives~\cite{kim2024stableviton,leffa}, leveraging reference networks~\cite{Zhu_2023_CVPR_tryondiffusion,idm-vton,xu2024ootdiffusion}, or injecting conditioning signals through spatial concatenation~\cite{chong2024catvtonconcatenationneedvirtual}.
Despite these efforts, existing methods often fail to establish precise garment--person correspondence. As illustrated in Fig.~\ref{fig:intro}, the dispersed attention patterns reflect limited relational understanding, often resulting in lower visual quality and failure to faithfully reconstruct garment details.

To address these challenges, we introduce \emph{\textbf{Voost}}, a unified framework that jointly learns virtual try-on and its inverse task, virtual try-off, which aims to reconstruct the original appearance of the worn garment from a person image, within a single diffusion transformer. \emph{Voost} adopts a token-level concatenation structure, where spatially aligned garment and person images are fed into a shared embedding space. This design enables the model to reason bidirectionally across try-on and try-off scenarios using a common conditioning layout. By jointly training both directions, \emph{Voost} strengthens garment--person correspondence while requiring no task-specific networks, auxiliary losses, or additional annotations, as each garment-person pair naturally provides supervision for the reverse process.

Our framework leverages token-level concatenation and the architectural flexibility of diffusion transformers to handle diverse poses, aspect ratios, and spatial layouts.
A task token encodes both generation direction and garment category, enabling scalable learning without task-specific models or fixed resolutions.
This unified setup not only supports multitask learning but also mitigates architectural, task-specific, and category-specific inductive biases by exposing the model to broader structural variation.

In addition, we propose two simple inference-time techniques to improve robustness. Attention temperature scaling mitigates resolution or mask size mismatch between training and test time. We also introduce self-corrective sampling, where try-on and try-off predictions iteratively refine each other within the unified model.

Remarkably, our unified diffusion transformer achieves state-of-the-art performance on both VTON and VTOFF benchmarks, surpassing strong task-specific baselines~\cite{kim2024stableviton,idm-vton,leffa,chong2024catvtonconcatenationneedvirtual,tryoffdiff,tryoffanyone} optimized individually for each task. We provide extensive qualitative and quantitative analyses that validate the effectiveness of our unified approach and its ability to model garment--person interaction across diverse scenarios.

\section{Related work}
\label{sec:related}

\paragraph{Image-conditioned diffusion models.}
Diffusion models have rapidly evolved to support image-conditioned generation beyond the text-to-image setting~\cite{rombach2022high,podell2024sdxl}. Early methods such as Textual Inversion~\cite{gal2023an} and DreamBooth~\cite{ruiz2023dreambooth} enabled concept-specific synthesis via embedding injection or fine-tuning, but incurred high computational costs.
Later approaches improved efficiency by leveraging pre-trained encoders like CLIP~\cite{radford2021learning} and DINO~\cite{caron2021emerging,oquab2024dinov}, as seen in IP-Adapter~\cite{ye2023ip-adapter} and AnyDoor~\cite{chen2024anydoor}, which use image embeddings to guide generation.
To improve spatial consistency, MasaCtrl~\cite{cao2023masactrl} introduced mutual self-attention to preserve object identity. This was extended by reference-based methods~\cite{Hu_2024_CVPR,chen2024zero,xu2024ootdiffusion,idm-vton,tian2024emo} that inject structured visual features through dedicated networks.
More recently, concatenation-based methods~\cite{chong2024catvtonconcatenationneedvirtual,lhhuang2024iclora,tan2024ominicontrol} have emerged as lightweight alternatives that directly feed conditioning signals without auxiliary modules.
Additionally, image-conditioned diffusion has expanded beyond image-to-image generation to video~\cite{guo2023animatediff,Hu_2024_CVPR,blattmann2023stable,shi2024motion} and 3D synthesis~\cite{liu2023zero,kwak2024vivid,voleti2024sv3d}, broadening its applicability.

\paragraph{Virtual try-on using generative models.}
Early virtual try-on systems primarily relied on warping-based methods~\cite{wang2018toward,han2019clothflow,ge2021parser}, where the main challenge was to deform and align the garment image to match the target person’s pose and body shape.
With the emergence of Generative Adversarial Networks (GANs)~\cite{goodfellow2014generative,karras2019style,karras2020analyzing}, a series of GAN-based approaches~\cite{Han_2018_CVPR,xie2022pastagan,Men_2020_CVPR_gan,issenhuth2020not,xie2023gpvton_gan,bai2022singlestagevirtualtryon_gan,ge2021parser_gan,Yang_2020_CVPR_gan,choi2021viton} achieved more realistic image synthesis through adversarial learning.
However, these models still struggled to preserve garment details and maintain structural consistency.

With the advancement of large-scale text-to-image diffusion models~\cite{ramesh2021zero,rombach2022high,saharia2022photorealistic,podell2024sdxl,esser2024scalingrectifiedflowtransformers,flux2024}, recent works~\cite{Zhu_2023_CVPR_tryondiffusion,baldrati2023multimodal,morelli2023ladi,gou2023taming_dci,idm-vton,xu2024ootdiffusion,zeng2024cat,shen2025imagdressing,chong2024catvtonconcatenationneedvirtual,leffa} have adopted pre-trained diffusion backbones as strong generative priors for virtual try-on.
A key challenge remains: effectively guiding the model to learn the complex relationship between the garment and the target person. To address this, various techniques have been proposed to improve spatial correspondence and visual fidelity—such as auxiliary loss functions~\cite{kim2024stableviton,leffa}, dual or reference networks~\cite{Zhu_2023_CVPR_tryondiffusion,idm-vton,chen2024wear}, warping~\cite{morelli2023ladi,gou2023taming_dci,xie2023gpvton_gan,leffa}, external image encoders~\cite{kim2024stableviton,idm-vton,xu2024ootdiffusion}, and spatial concatenation~\cite{chong2024catvtonconcatenationneedvirtual}.
However, many of these approaches rely on weak garment–person coupling or introduce additional components that may degrade generation quality.

While most existing work focuses on virtual try-on, a few studies~\cite{tryoffdiff,tryoffanyone} have explored virtual try-off, the inverse of virtual try-on, which aims to predict a clean garment image from a person wearing it. This task is inherently challenging due to occlusion, wrinkles, and deformation, and similarly requires strong relational modeling.
We propose a unified DiT-based framework that jointly learns virtual try-on and try-off using a shared spatial conditioning layout. This design enables robust bidirectional garment–person modeling and remains scalable across diverse fashion imagery.

\begin{figure*}[!t]
    \centering
    \includegraphics[width=\linewidth]{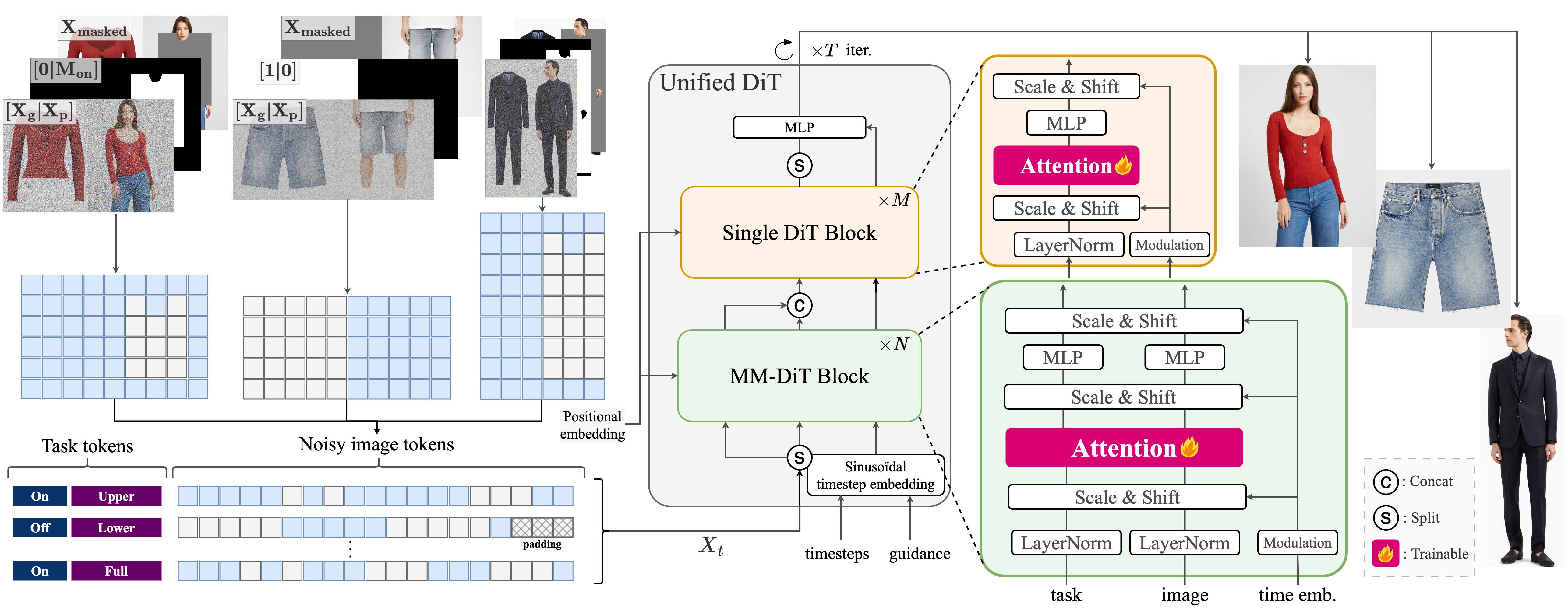}
    \caption{
Overview of pipeline. \emph{Voost} enables bidirectional virtual try-on and try-off with a unified diffusion transformer for scalable learning.
}
    \label{fig:method}
\end{figure*}

\section{Method}
\label{sec:method}
In this section, we introduce our framework for virtual try-on and try-off using a unified diffusion transformer. We first present the preliminaries of our approach (Sec.~\ref{subsec:prel}), and then describe how we design a diffusion architecture that supports both tasks within a shared model (Sec.~\ref{subsec:pipeline}). Finally, we introduce inference-time techniques that further enhance output quality (Sec.~\ref{subsec:inference}).

\subsection{Preliminary} \label{subsec:prel}
We begin by reviewing conditional diffusion models in the context of virtual try-on, and motivating the need for strong garment–target correspondence to enable faithful synthesis.

\paragraph{Diffusion models for virtual try-on.} 
Diffusion models~\cite{ho2020denoising,song2021scorebased,rombach2022high} generate images by denoising a sequence of latent representations corrupted over time. 
Given a latent \(\mathbf{z}_t\), the network \(\epsilon_\theta\) is trained to recover the clean signal \(\mathbf{z}_0\) or the noise \(\boldsymbol{\epsilon}\), conditioned on external input \(\mathbf{y}\). 
In general, the denoising step follows:
\begin{equation}
\mathbf{z}_{t-1} = \sample\left(\mathbf{z}_t, \epsilon_\theta(\mathbf{z}_t, t, \mathbf{y})\right),
\end{equation}
where \(\sample(\cdot)\) denotes an integration step, such as DDIM~\cite{song2021denoising}, DPM-Solver~\cite{lu2022dpm}, or deterministic alternatives like flow matching~\cite{lipman2023flowmatchinggenerativemodeling}.  
In virtual try-on, \(\mathbf{y}\) typically includes a garment image and a masked person image, optionally with pose or parsing maps~\cite{cao2017realtime,guler2018densepose}.  
We adopt a unified DiT~\cite{Peebles2022DiT} that handles both try-on and try-off tasks. Details follow in Sec.~\ref{subsec:pipeline}.

\paragraph{Garment–target correspondence analysis.}
A key challenge in virtual try-on is aligning the garment structure with the target body while preserving visual fidelity.
We assess this correspondence by analyzing transformer attention maps at specific spatial queries.
As shown in Fig.\ref{fig:intro}, CatVTON~\cite{chong2024catvtonconcatenationneedvirtual} exhibits dispersed attention across unrelated areas, indicating weak spatial grounding, which can reduce structural fidelity and impair detail preservation.  
In contrast, attention maps produced by our model are sharply localized and semantically consistent, suggesting stronger garment--target alignment.  
This motivates our bidirectional training strategy, where jointly learning try-on and try-off within a single model improves spatial alignment.

\subsection{Try-On \& Off via Unified Transformer} \label{subsec:pipeline}

As shown in Fig.~\ref{fig:method}, we use a horizontally concatenated layout that places the garment image and the person image side by side, allowing the model to process both conditioning and generation regions within a unified input.
While this setup enables joint processing of conditioning and generation regions, existing methods often reconstruct the conditioning region without masking, turning it into a trivial task.
To address this, we extend the setup for bidirectional learning: by selectively masking the garment region while showing the dressed person, we reinterpret the task as virtual try-off, allowing the model to infer the garment from contextual cues.

\paragraph{Pipeline.}
Let $\mathbf{X_g} \in \mathbb{R}^{H \times W \times 3}$ denote the standalone garment image, and $\mathbf{X_p} \in \mathbb{R}^{H \times W \times 3}$ the target person image.
We construct the input by horizontally concatenating the two, forming a combined image $\mathbf{X} = [\mathbf{X_g} \mid \mathbf{X_p}] \in \mathbb{R}^{H \times 2W \times 3}$.
Task-specific inpainting regions are defined by a binary mask $\mathbf{M} \in \{0,1\}^{H \times 2W}$, applied in the image space prior to encoding. 
For the try-on task, the mask is defined as $\mathbf{M} = [\mathbf{0} \mid \mathbf{M_{\text{on}}}]$, where $\mathbf{M_{\text{on}}}$ masks out the garment area in the person image $\mathbf{X_p}$ while preserving the garment image $\mathbf{X_g}$.
For the try-off task, we use $\mathbf{M} = [\mathbf{1} \mid \mathbf{0}]$, masking the entire garment image and leaving the person image unmasked.
The masked input is obtained as $\mathbf{X_{\text{masked}}} = \mathbf{X} \odot (1 - \mathbf{M})$, where $\odot$ denotes element-wise multiplication.
Our architecture follows a latent diffusion framework~\cite{flux2024,esser2024scalingrectifiedflowtransformers}, where denoising is performed in the latent space.  

The full and masked images are encoded by a frozen encoder $\mathcal{E}$ into latent representations $\mathbf{z}_0 = \mathcal{E}(\mathbf{X})$ and $\mathbf{z}_{\text{c}} = \mathcal{E}(\mathbf{X_{\text{masked}}})$.  
The mask $\mathbf{M}$ is downsampled via a pixel-unshuffle~\cite{shi2016real} operation to match the latent resolution, producing $\mathbf{M}_{\text{c}}$.
These three components $\mathbf{z}_0, \mathbf{z}_{\text{c}}$, and $ \mathbf{M}_{\text{c}}$ are concatenated along the channel dimension, passed through a token embedding layer, and then used as input to the DiT backbone.
We define the task token $\tau = [\tau_{\text{mode}} \mid \tau_{\text{category}}]$ as the concatenation of a mode token $\tau_{\text{mode}} \in \{\text{on}, \text{off}\}$, which specifies the generation direction, and a category token $\tau_{\text{category}} \in \{\text{upper}, \text
{lower}, \text{full}\}$, which encodes the garment type. The combined token $\tau$ is passed to the transformer as additional conditioning.

\paragraph{Dynamic layout.}
Our framework supports dynamic input layouts by leveraging the token-based representation of vision transformers~\cite{dosovitskiy2021an_vit,dehghani2023patch,Peebles2022DiT}, which enables the processing of images with varying spatial resolutions and aspect ratios without requiring fixed dimensions.
In our pipeline, the concatenated garment–person image is tokenized and flattened, allowing the model to accommodate inputs of arbitrary shape. Consequently, the aforementioned $H$ and $W$ need not remain fixed throughout the pipeline.
To support batch training with variable-length sequences, we insert padding tokens as needed and standardize the sequence length to a fixed maximum \( N_{\max} \).
Additionally, we also use Rotary Position Embedding (RoPE)~\cite{su2024roformer} to robustly handle inputs with diverse aspect ratios and spatial configurations.
This design provides the foundation for scalable virtual try-on, enabling \emph{Voost} to handle diverse poses, aspect ratios, and compositions within a single batch.
By jointly modeling task direction and garment category, it supports flexible multitask learning and mitigates inductive biases introduced by the rigid preprocessing of fixed-resolution pipelines.

\paragraph{Training strategy.}
We adopt a flow matching formulation~\cite{lipman2023flowmatchinggenerativemodeling}, where the model learns a time-dependent velocity field that transports samples from the data distribution to noise along a continuous path.
Let \( \mathbf{z}_0 \) be a data latent and \( \mathbf{z}_1 \sim \mathcal{N}(\mathbf{0}, \mathbf{I}) \) be a sampled noise latent.
We define a trajectory \( \mathbf{z}_t \) between them and train the denoising model \( \epsilon(\cdot) \) to predict the velocity \( \frac{d\mathbf{z}_t}{dt} \) at each intermediate point.  
The unified training objective is given by:
\begin{equation}
\mathcal{L}_{\text{unified}} = \mathbb{E}_{t,\, \mathbf{z}_0,\, \mathbf{z}_1} \left[ \left\| \epsilon(\mathbf{z}_t, \mathbf{z}_{\text{c}}, \mathbf{M}, \tau, t) - \frac{d\mathbf{z}_t}{dt} \right\|^2 \right].
\end{equation}

In our case, we adopt the rectified flow formulation~\cite{liu2022flow,esser2024scalingrectifiedflowtransformers}, where the trajectory is defined as a straight line between $\mathbf{z}_0$ and $\mathbf{z}_1$:
\begin{equation}
\mathbf{z}_t = (1 - t)\mathbf{z}_0 + t\mathbf{z}_1, \quad \text{where } t \in [0, 1].
\end{equation}
Note that $t$ denotes continuous time during training, but serves as a discrete timestep index during inference.
This reduces the training target to a constant displacement vector, simplifying optimization and aligning with the rectified flow formulation.
To retain the strong generative prior of the pretrained DiT backbone while adapting to the try-on and try-off setting, we finetune only the attention modules within each transformer block, freezing the rest.
Adapting only attention layers enables precise garment–person reasoning in try-on and try-off, without sacrificing the generative prior learned from large-scale diffusion.
A detailed analysis of this design choice is provided in the \textit{supplementary material}.

\begin{figure}[t!]
    \centering
    \includegraphics[width=\linewidth]{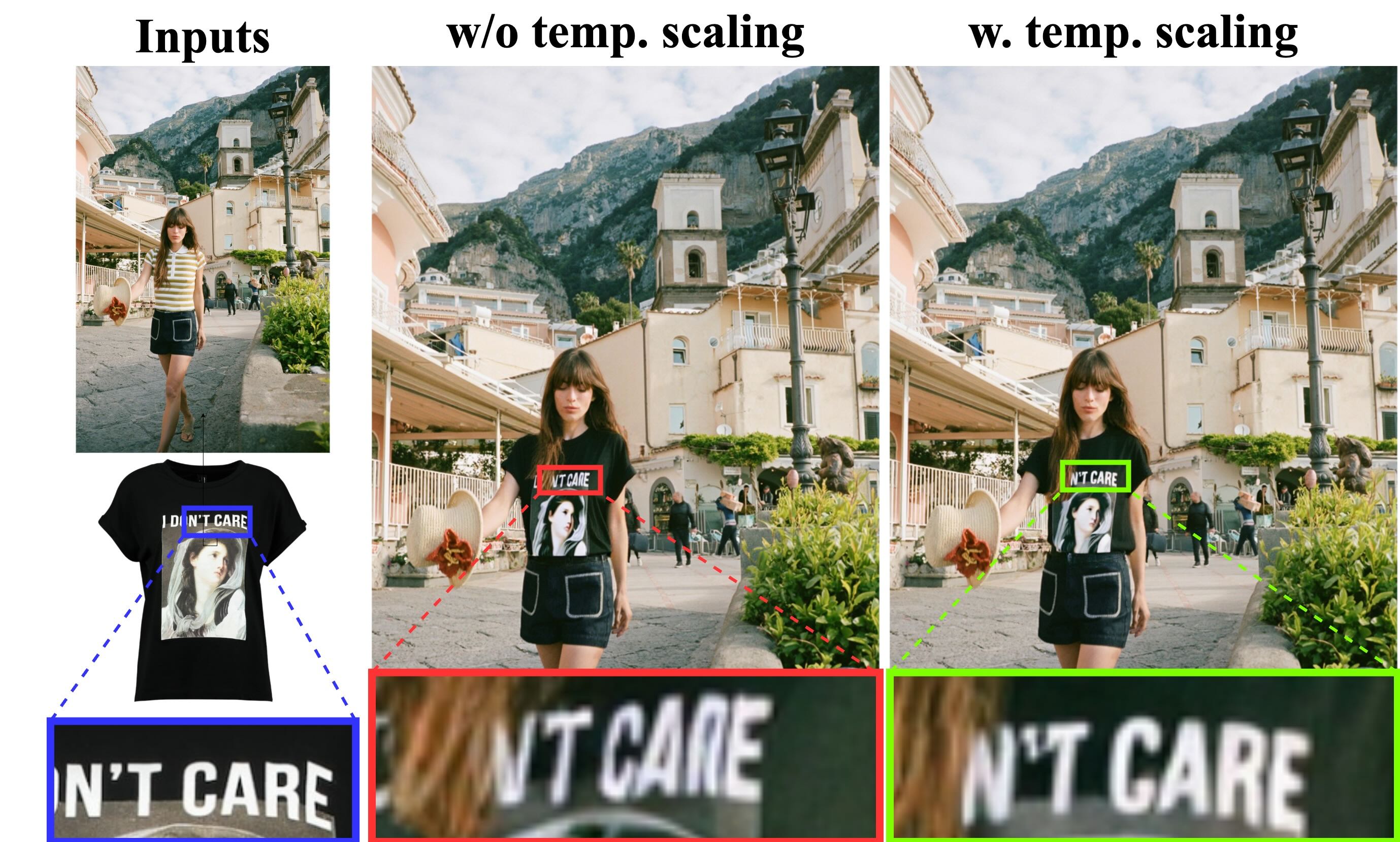}
\caption{
Impact of temperature scaling.
Adaptive temperature scaling enhances visual detail by adjusting attention behavior under varying spatial proportions of mask and garment regions.
}
    \label{fig:inference_temp_scaling} 
\end{figure}

\subsection{Inference-time Refinement} \label{subsec: inference}

\paragraph{Token-length and mask-aware temperature scaling.} 
While our framework flexibly supports varying spatial aspect ratios during training, real-world inference often involves inputs with resolutions or mask proportions that differ from those seen during training. 
Such shifts can lead to suboptimal attention behavior, especially when inputs fall outside the training distribution. To address this, we introduce a dynamic temperature scaling scheme that adapts attention sharpness to both absolute token length changes and relative spatial imbalance between the masked and conditioning regions. We define the adjusted attention temperature as:

\begin{equation}
\lambda' = 
\underbrace{ \sqrt{ \frac{1}{d} } }_{\scriptsize \smash{ \text{base scaling} } } 
\cdot 
\underbrace{ \sqrt{ \alpha \cdot \frac{ \log \! \left( N_{\text{infer}} \right) }{ \log \! \left( N_{\text{train}} \right) } } }_{\scriptsize \smash{ \text{global token scaling} } } 
\cdot 
\underbrace{ \sqrt{ \frac{ \log \left( N_{\text{mask}} + c \right) }{ \log \left( \beta \cdot N_{\text{garment}} + c \right) } } }_{\scriptsize \smash{ \text{mask-aware relative scaling} } }.
\end{equation}

\sy{

}

Here, \(d\) is the key, query, value vector dimension used in scaled dot-product attention~\cite{vaswani2017attention}; \(N_{\text{infer}}\) and \(N_{\text{train}}\) are the total token counts at inference and training; \(N_{\text{mask}}\) and \(N_{\text{garment}}\) denote the masked and garment region tokens.
The hyperparameters \(\alpha\) and \(\beta\) control the influence of the global and relative scaling terms, respectively, while \(c\) is a fixed positive constant that stabilizes the computation when \(N_{\text{mask}}\) is small.
The global term stabilizes attention across token lengths, while the relative term adapts to spatial imbalance between masked and garment regions.
This improves robustness to layout shifts and maintains consistent attention, particularly when the mask covers a small proportion relative to the garment (Fig.~\ref{fig:inference_temp_scaling}).

\begin{algorithm}[!t]
\caption{Self-corrective sampling }
\label{alg:self-correction}
\KwIn{Noisy latent $\mathbf{z}_T$, garment latent $\mathbf{z}_{\text{c}}$, target mask $\mathbf{M}_{\text{on}}$, denoiser $\epsilon_\theta$, correction timestep set $\mathcal{T}_{\text{corr}}$, refinement iteration $R$, noise schedule $\{ \sigma_t \}_{t=1}^T$}
\KwOut{Final latent $\hat{\mathbf{z}}_0$}

\For{$t = T, \dots, 1$}{
  $\mathbf{M} \gets [\mathbf{0} \mid \mathbf{M}_{\text{on}}]$\;
  $\hat{\mathbf{v}}^{\text{on}}_t = \epsilon(\mathbf{z}_t, \mathbf{X}_{\mathrm{g}}, \mathbf{M}, \tau_{\text{on}}, t)$ \;
  
$\mathbf{z}_{t-1} = \mathbf{z}_t + (\sigma_{t-1} - \sigma_t) \cdot \hat{\mathbf{v}}_t$
  
  \If{$t \in \mathcal{T}_{\text{corr}}$}{
    $\hat{\mathbf{z}}^{\text{on}}_0 = \texttt{predict\_x0}(\mathbf{z}_t, \hat{\mathbf{v}}^{\text{on}}_t)$\;

    $\mathbf{M} \gets [\mathbf{1} \mid \mathbf{0}]$\;
    $\hat{\mathbf{v}}^{\text{off}}_t = \epsilon(\mathbf{z}_t, \hat{\mathbf{z}}^{\text{on}}_0, \mathbf{M}, \tau_{\text{off}}, t)$\;
    $\hat{\mathbf{z}}^{\text{off}}_0 = \texttt{predict\_x0}(\mathbf{z}_t, \hat{\mathbf{v}}^{\text{off}}_t)$\;

    \For{$r = 1, \dots, R$}{
      $\mathbf{z}_t \gets \mathbf{z}_t - \eta \cdot \nabla_{\mathbf{z}_t} \left\| \hat{\mathbf{z}}^{\text{off}}_0 - \mathbf{z}_{\text{c}} \right\|^2$\;

    }
  }
}
\Return{$\hat{\mathbf{z}}_0$}
\end{algorithm}
\paragraph{Self-correction with unified transformer.} \label{subsec:inference}
Our unified model enables both try-on and try-off generation under a shared concatenated layout. 
We leverage this dual capability at inference time through a \textit{self-correction mechanism} (Alg.~\ref{alg:self-correction}) that promotes consistency between the generated output and the conditioning garment.
The core intuition is that a faithful try-on result should implicitly preserve sufficient information to recover the original garment via a reverse (try-off) process. 
 At a designated timestep $t \in \mathcal{T}_{\text{corr}}$ during denoising, the model predicts the dressed person image $\hat{\mathbf{z}}^{\text{on}}_0$ from the current latent $\mathbf{z}_t$, using the try-on task token $\tau_{\text{on}}$ (i.e., the mode component of the full task token $\tau = [\tau_{\text{mode}} \mid \tau_{\text{category}}]$). 
 This intermediate prediction is then used as a conditioning input to perform a reverse try-off pass, yielding a reconstructed garment $\hat{\mathbf{z}}^{\text{off}}_0$.
We compute reconstruction error between $\hat{\mathbf{z}}^{\text{off}}_0$ and $\mathbf{z}_{\text{c}}$, using its gradient to update $\mathbf{z}_t$ via backpropagation. This refinement is repeated $R$ times to gradually align the generation with the conditioning signal.

\begin{figure*}[!t]
    \centering
    \includegraphics[width=\linewidth]{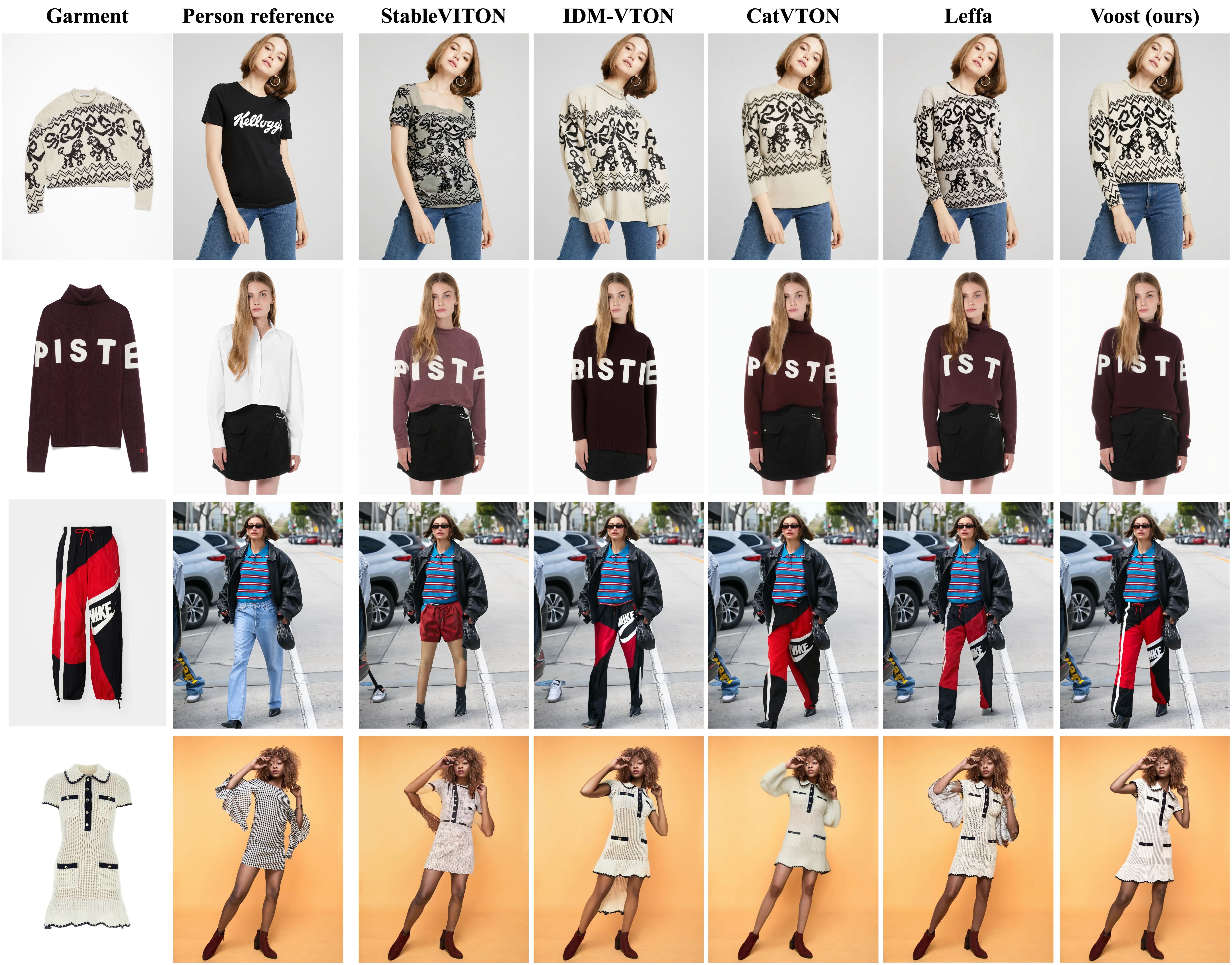}
    \caption{Qualitative comparison of try-on results with existing try-on methods~\cite{kim2024stableviton,idm-vton,chong2024catvtonconcatenationneedvirtual,leffa}.
    Best viewed in color and under zoom.}
    \label{fig:qualitative}
\end{figure*}

\section{Experiments}
\label{sec:result}
\subsection{Datasets and experimental setup}

\begin{figure*}[!t]
    \centering
    \includegraphics[width=\linewidth]{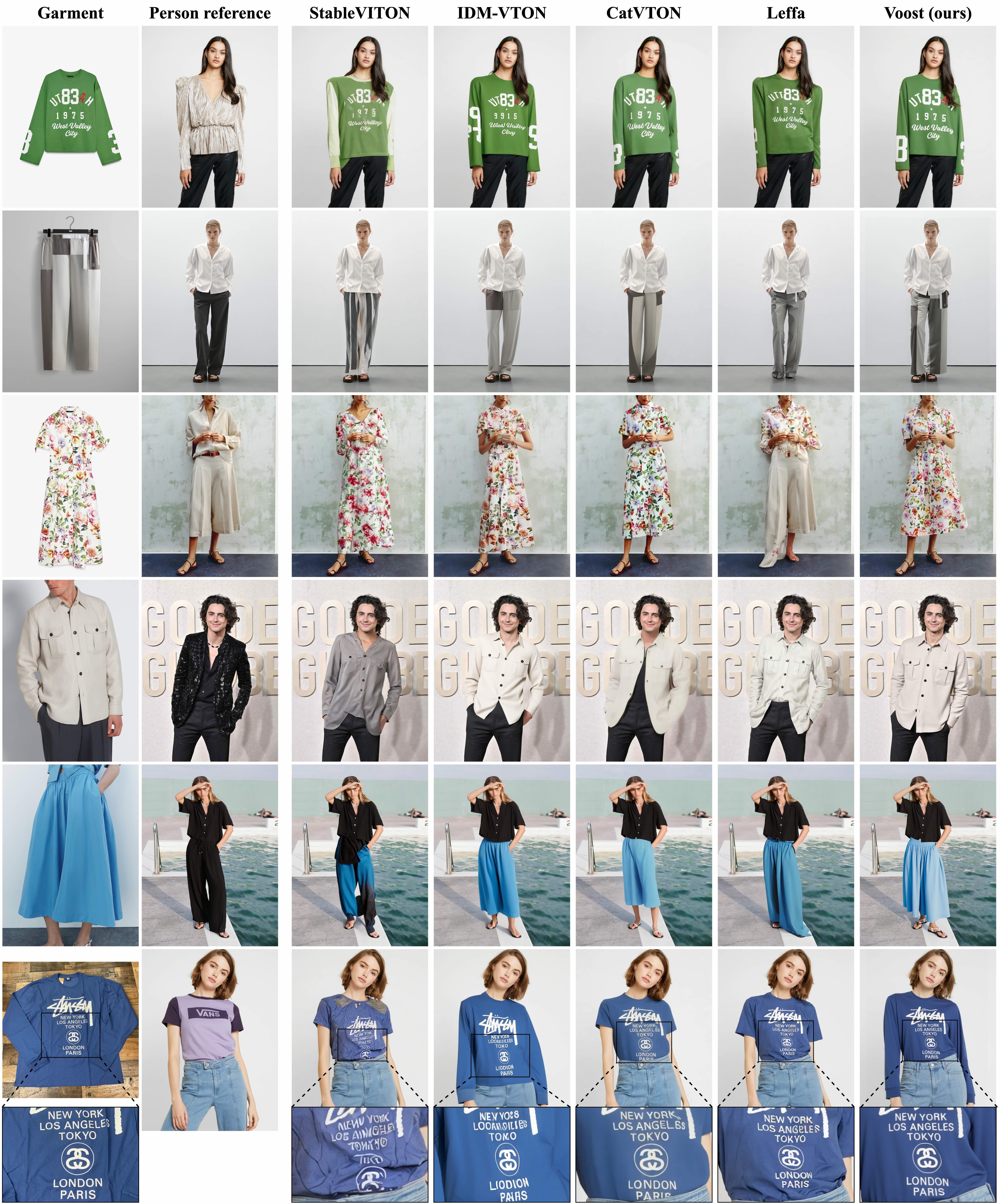}

    \caption{Additional qualitative comparison of try-on results with other methods.
    Best viewed in color and under zoom.}
    \label{fig:fig_only_qual_tryon} 
\end{figure*}

\begin{figure}[!t]
    \centering
    \includegraphics[width=\linewidth]{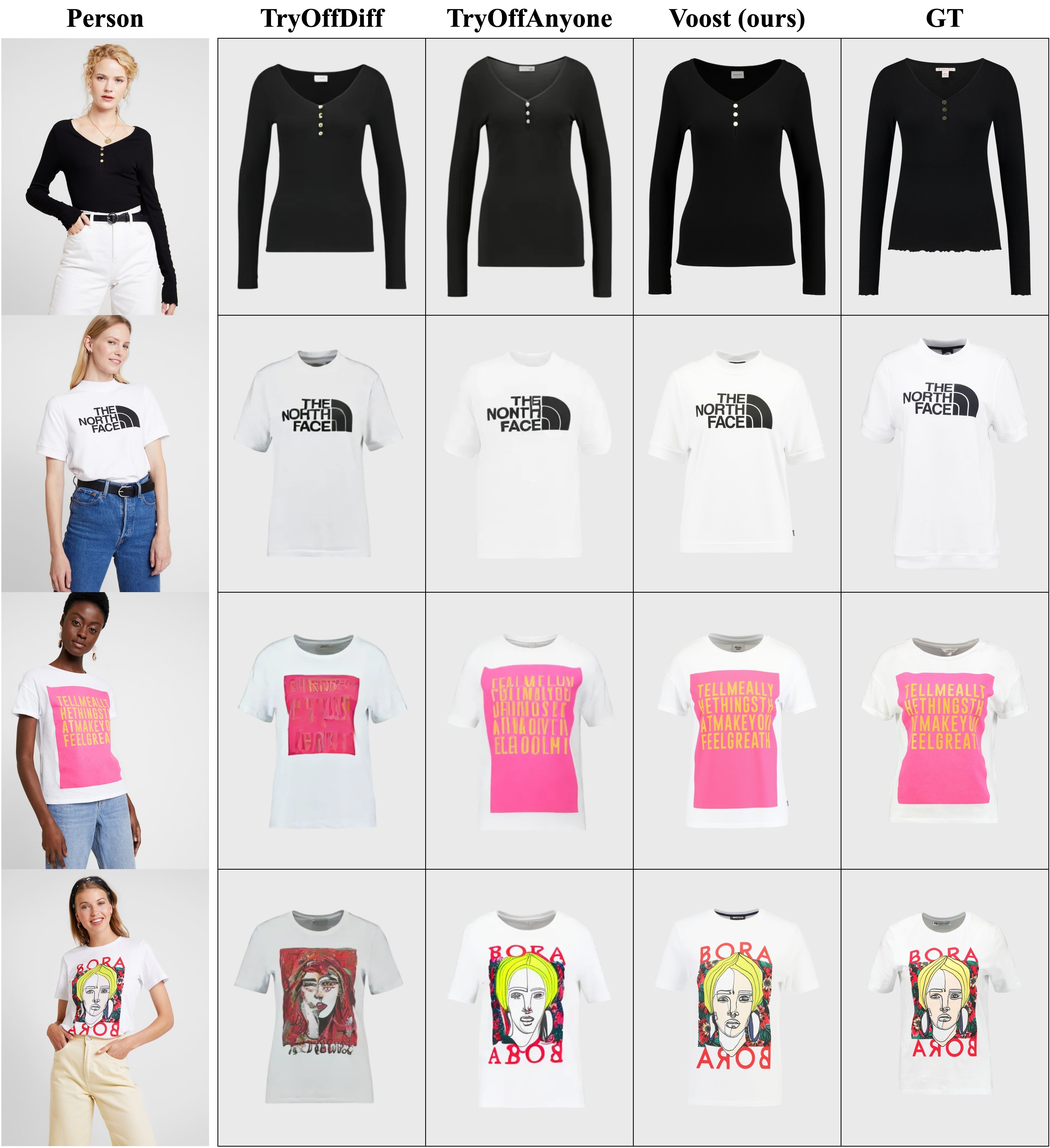}
    \caption{Qualitative comparison of try-off results with other methods.
    Best viewed in color and under zoom.}
    \label{fig:qual_tryoff}
\end{figure}
 
\begin{figure}[!t]
    \centering
    \includegraphics[width=0.98\linewidth]{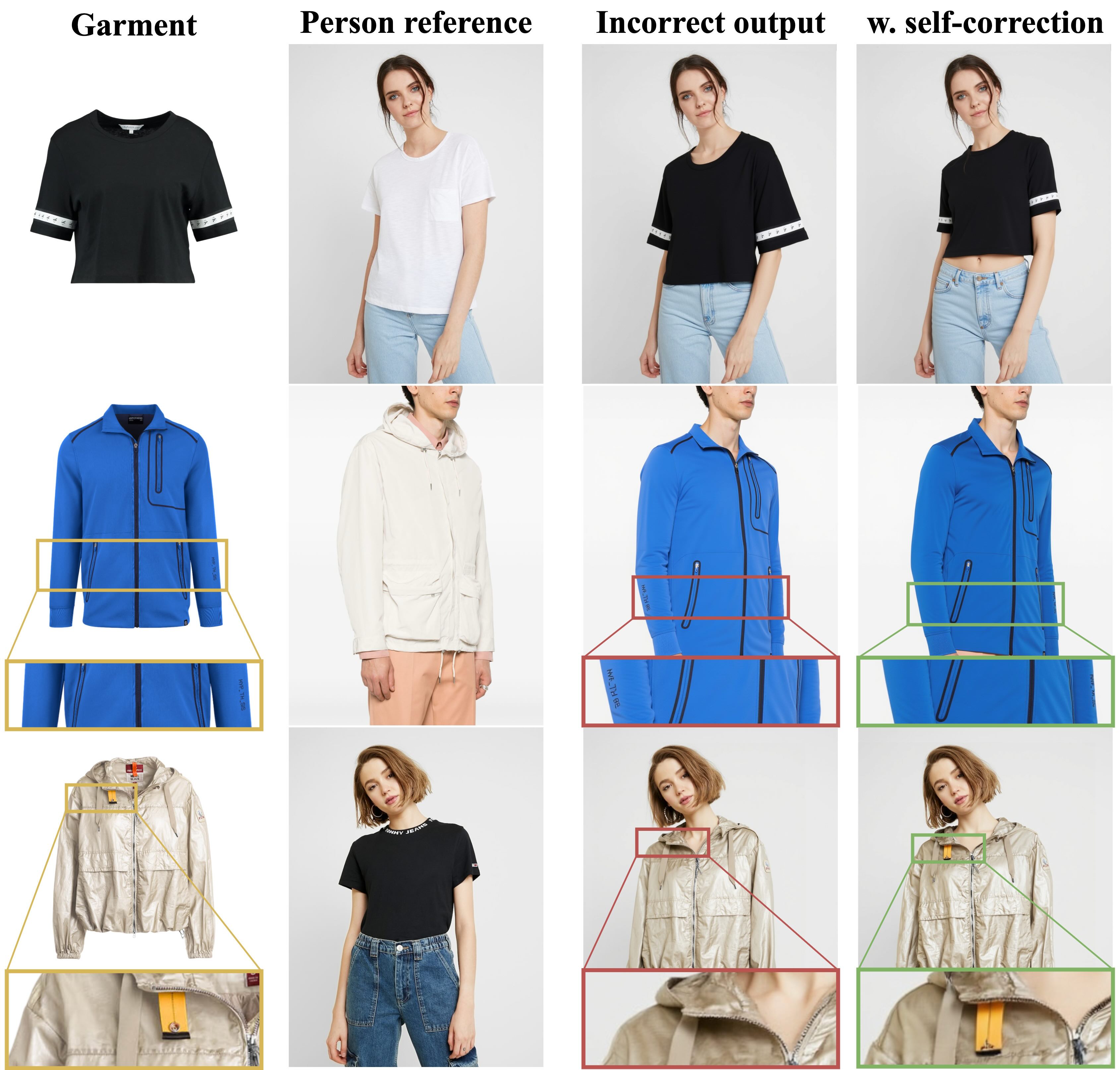}
    \caption{Effectiveness of the self-correction mechanism. These examples demonstrate improved garment–body alignment and reduction of visual artifacts when the self-correction module is applied.}
    \label{fig:inference_self_correction} 
\end{figure}

We evaluate our method on two standard benchmarks: DressCode~\cite{morelli2022dress} and VITON-HD~\cite{choi2021viton} datasets, using both qualitative and quantitative metrics.
Each dataset contains high-resolution image pairs of in-shop garments and corresponding person images.
We use the official train/validation/test splits provided by both datasets.
All outputs are generated at a resolution of $1024 \times 768$.
Our model is trained on $16 \times$ NVIDIA H100 GPUs and evaluated on a single NVIDIA A100 GPU.
To assess generalization, we also present qualitative results on in-the-wild images.

\subsection{Qualitative comparison}

\begin{figure*}[!t]
    \centering
     \includegraphics[width=\linewidth]{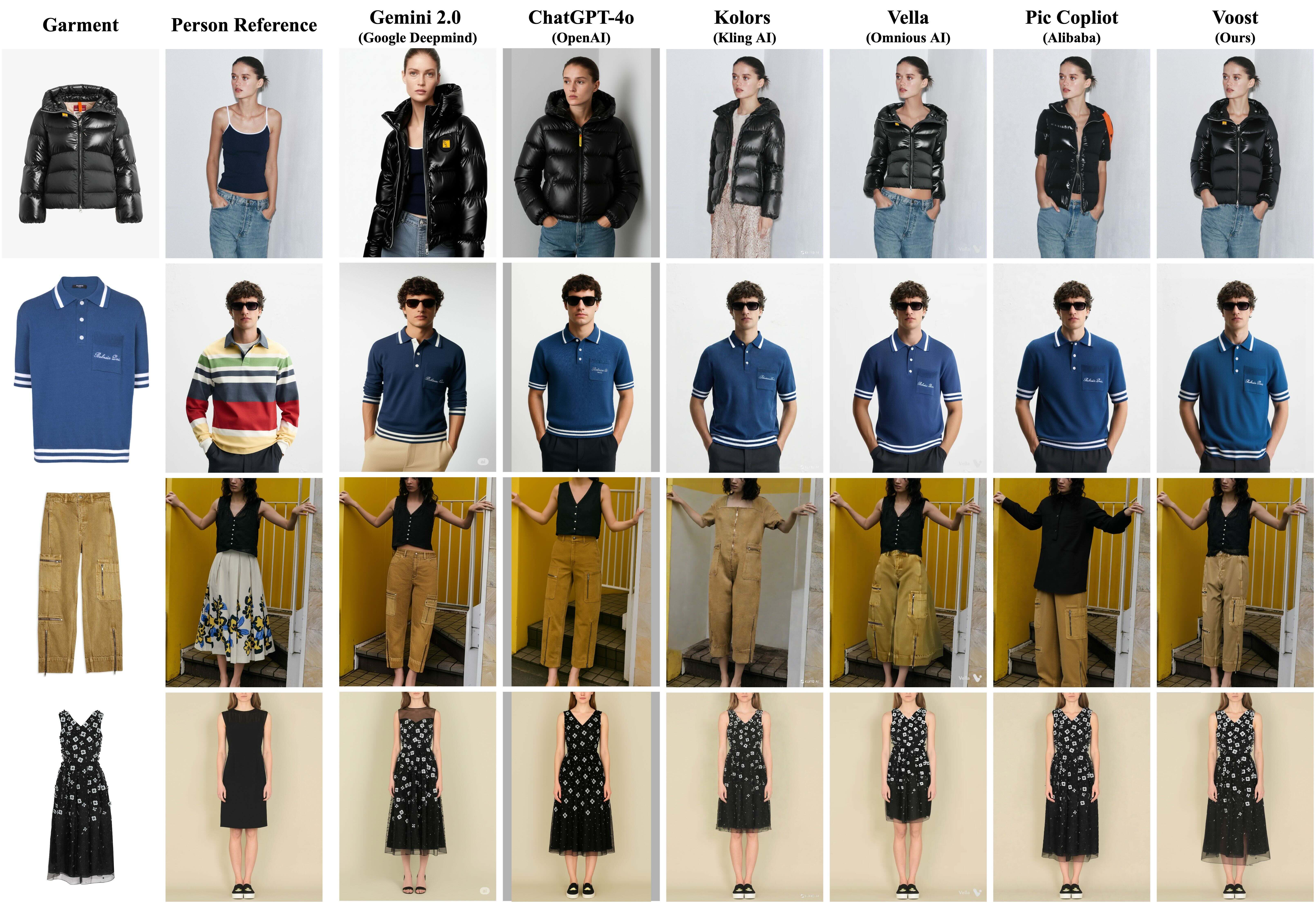}
    \caption{Qualitative comparison of try-on results with other commercial models ~\cite{gemini2, chatgpt4o, kolors, vella, piccopilot}. Best viewed in color and under zoom.}
    \label{fig:fig_only_qual_tryon_commercial} \vspace{-2mm}
\end{figure*}

Fig.~\ref{fig:qualitative} and Fig.~\ref{fig:fig_only_qual_tryon} show qualitative comparisons between our method and state-of-the-art approaches~\cite{kim2024stableviton, idm-vton, chong2024catvtonconcatenationneedvirtual, leffa} on the VTON task. Given the increasing interest in practical applications of virtual try-on, we also include comparisons with commercial models~\cite{gemini2,chatgpt4o,kolors,vella,piccopilot} in Fig.~\ref{fig:fig_only_qual_tryon_commercial}. 

\begin{figure}[!t]
    \centering
    \includegraphics[width=\linewidth]{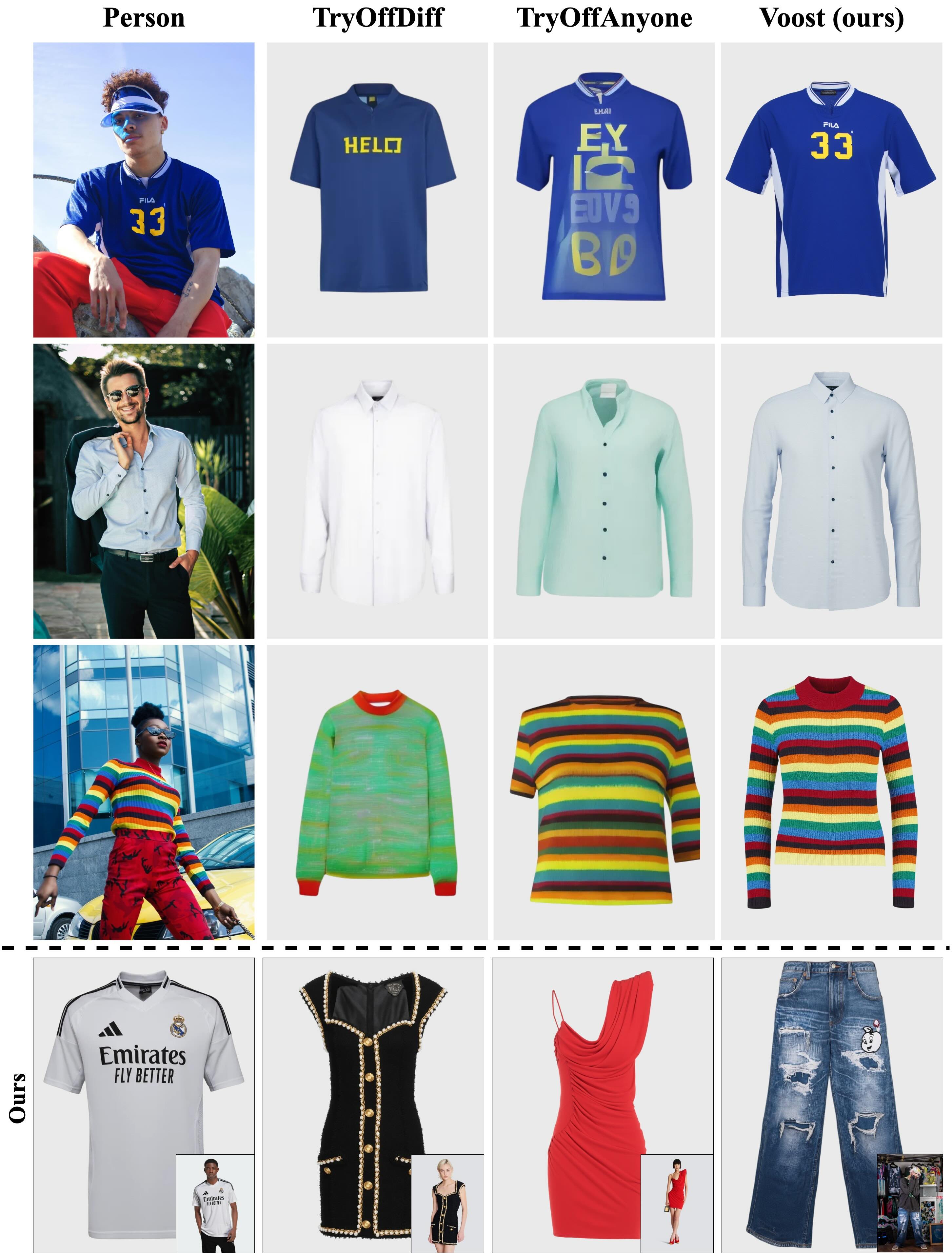}
\caption{
Comparison results on in-the-wild images (Row 1–3) for the virtual try-off task. 
Unlike existing methods~\cite{tryoffdiff,tryoffanyone}, our method supports diverse garment types including upper, lower, and dresses (Last row).
}
    \label{fig:fig_only_tryoff} 
\end{figure}

Importantly, our unified model handles try-on and try-off with a single set of parameters, without task-specific retraining. 
Try-off results are shown in Fig.~\ref{fig:qual_tryoff} and Fig.~\ref{fig:fig_only_tryoff}, where we compare our outputs to those of existing methods~\cite{tryoffdiff, tryoffanyone}. 
Across both tasks, our method produces more coherent and photorealistic results, benefiting from shared garment-body reasoning and joint training.

In particular, our model demonstrates strong robustness on in-the-wild images with diverse poses, backgrounds, and lighting conditions, consistently outperforming existing methods in these challenging scenarios.
We provide additional qualitative results and comparisons in the \textit{supplementary material}.

\subsection{Quantitative results}

\begin{table*}[!t]
    \footnotesize
    \centering
    \setlength{\tabcolsep}{8pt}
    \begin{tabular}{@{}lcccccccccccc@{}}
        \toprule
        \textbf{Methods} & 
        \multicolumn{6}{c}{\textbf{VITON-HD~\cite{choi2021viton}}} & 
        \multicolumn{6}{c}{\textbf{DressCode~\cite{morelli2022dress}}} \\
        & \multicolumn{4}{c}{Paired} & \multicolumn{2}{c}{Unpaired} & \multicolumn{4}{c}{Paired} & \multicolumn{2}{c}{Unpaired} \\
        \cmidrule(lr){2-5} \cmidrule(lr){6-7} \cmidrule(lr){8-11} \cmidrule(lr){12-13}
        & SSIM↑ & LPIPS↓ & FID↓ & KID↓ & FID↓ & KID↓ & SSIM↑ & LPIPS↓ & FID↓ & KID↓ & FID↓ & KID↓ \\
        \midrule
        StableVITON~\cite{kim2024stableviton} & 0.867 & 0.084 & 6.851 & 1.255 & 9.591 & 1.451 & 0.905 & 0.107 & 4.482 & 1.530 & 6.728 & 1.742 \\
        OOTDiffusion~\cite{xu2024ootdiffusion} & 0.851 & 0.096 & 6.520 & 0.896 & 9.672 & 1.206 & 0.898 & 0.073 & 3.953 & 0.720 & 6.704 & 1.863 \\
        IDM-VTON~\cite{idm-vton} & 0.881 & 0.079 & 6.343 & 1.322 & 9.613 & 1.639 &0.923 & 0.048 & 3.801 & 1.201 & 5.621 & 1.554 \\
        CatVTON~\cite{chong2024catvtonconcatenationneedvirtual} & 0.869 & 0.097 & 6.141 & 0.964 & 9.141 & 1.267 & 0.901 & 0.071 & 3.283 & 0.670 & 5.424 & 1.549 \\
        Leffa~\cite{leffa} & 0.872 & 0.081 & 6.310 & 1.208 & 9.442 & 1.249 & 0.911 & 0.060 & 3.651 & 0.709 & 5.462 & 1.528 \\
        \midrule
        \textbf{Ours (VTON-only)} & 0.868 & 0.079 & 5.804 & 0.618 & 9.112 & 1.051 & 0.910 & 0.052 & 3.043 & 0.565 & 5.298 & 1.132 \\
        \midrule
        \textbf{Ours (w/o $\lambda$ scaling)} & \underline{0.885} & \underline{0.072} & \textbf{5.242} & \underline{0.460} & \underline{8.991} & \underline{0.912} & \underline{0.925} & \underline{0.046} & \textbf{2.774} & \underline{0.390} & \textbf{5.030} & \underline{0.808} \\
        \textbf{Ours} & \textbf{0.898} & \textbf{0.056} & \underline{5.269} & \textbf{0.404} & \textbf{8.982} & \textbf{0.899} & \textbf{0.933} & \textbf{0.044} & \underline{2.787} & \textbf{0.377} & \underline{5.081} & \textbf{0.787} \\
        \bottomrule
    \end{tabular}
    \caption{Quantitative results on VITON-HD~\cite{choi2021viton} and DressCode~\cite{morelli2022dress} for the try-on task.
    We report both paired and unpaired evaluation results across benchmarks. 
    Our unified dual-task model (\emph{Voost}) consistently outperforms all baselines, including the single-task (VTON-only) model.
    \textbf{Bold} and \underline{underline} indicate the best and second-best scores, respectively.}
    \label{tab:quantitative_tryon} 
\end{table*}

We evaluate visual fidelity and structural consistency using standard metrics.
For realism, we report Fréchet Inception Distance (FID)~\cite{heusel2017gans} and Kernel Inception Distance (KID)~\cite{binkowski2018demystifying}.
To assess structural consistency, we use LPIPS~\cite{zhang2018unreasonableeffectivenessdeepfeatures} and SSIM~\cite{wang2004image}.
As shown in Table.~\ref{tab:quantitative_tryon} and Table.~\ref{tab:quantitative_tryoff}, our model outperforms existing methods across all metrics.

\subsection{User study}
While virtual try-on allows multiple plausible outputs, preserving garment characteristics and structure remains crucial for realism. However, standard metrics only partially capture key aspects of try-on quality, including garment realism and alignment consistency. 
To better evaluate perceptual fidelity, we conducted a user study comparing our method with existing approaches.
Participants evaluated 50 samples randomly selected from DressCode, VITON-HD, and in-the-wild images, judging three aspects: photorealism, garment detail, and garment structure. For each sample, they selected the most compelling result in each category. Full details on the criteria, survey questions, and interface are provided in the \textit{supplementary material}.
As shown in Fig.~\ref{fig:exp_user_study}, our method was consistently preferred across all evaluation criteria, demonstrating its superiority in visual realism and garment preservation.

\subsection{Ablation study}
\label{sec:ablation}

\begin{figure}[!t]
    \centering
    \includegraphics[width=\linewidth]{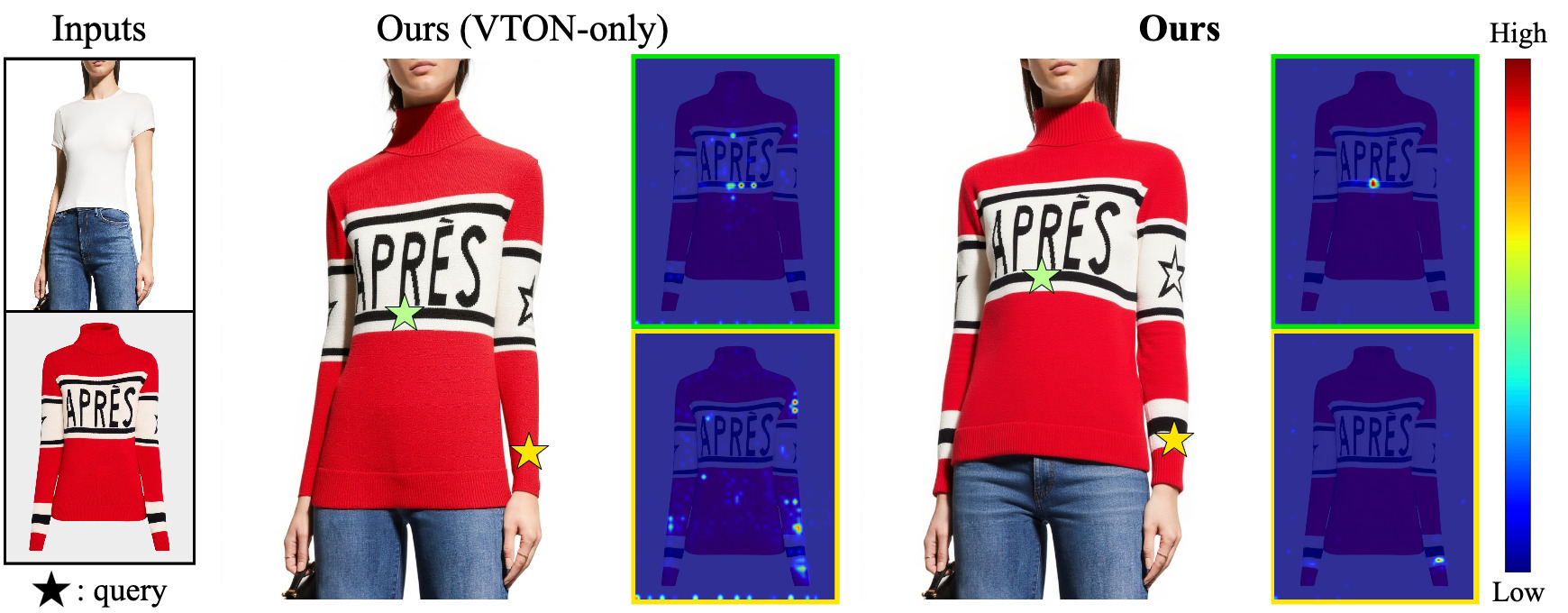}
    \caption{Attention map comparison of single-task and dual-task models. \emph{Voost} (dual-task) attends more precisely to the relevant garment region based on the query point.}
    \label{fig:exp_attn_maps}
\end{figure}

\paragraph{Effect of dual-task training.}
We examine the benefit of jointly training a single model for both try-on and try-off, compared to training separate models for each task.
As shown in Table.~\ref{tab:quantitative_tryon} and Table.~\ref{tab:quantitative_tryoff}, our dual-task model consistently outperforms its single-task counterparts in all metrics.
This indicates that joint training fosters a more generalizable garment–person correspondence prior, improving performance in both directions.
As illustrated in Fig.~\ref{fig:exp_attn_maps}, the attention maps of the joint model show sharper and more accurate garment-to-person correspondences, with regions in the person attending more precisely to their counterparts in the garment, highlighting improved spatial alignment.


\begin{table}[t]
    \centering
    \scriptsize
    \begin{tabular}{@{}l c c c c@{}}
        \toprule
         & TryOffDiff~\cite{tryoffdiff} & TryOffAnyOne~\cite{tryoffanyone} & \textbf{Ours (VTOFF-only)} & \textbf{Ours} \\
        \midrule
        FID$\downarrow$ & 28.25 & 25.20 & 12.88 & \textbf{10.06} \\
        KID$\downarrow$ & 11.42 & 6.98 & 3.54 & \textbf{2.48} \\
        \bottomrule
    \end{tabular}
    \caption{Quantitative results on VITON-HD for the try-off task.}
    \label{tab:quantitative_tryoff} 
\end{table}
\begin{figure}[!t]
    \centering
    \includegraphics[width=\linewidth]{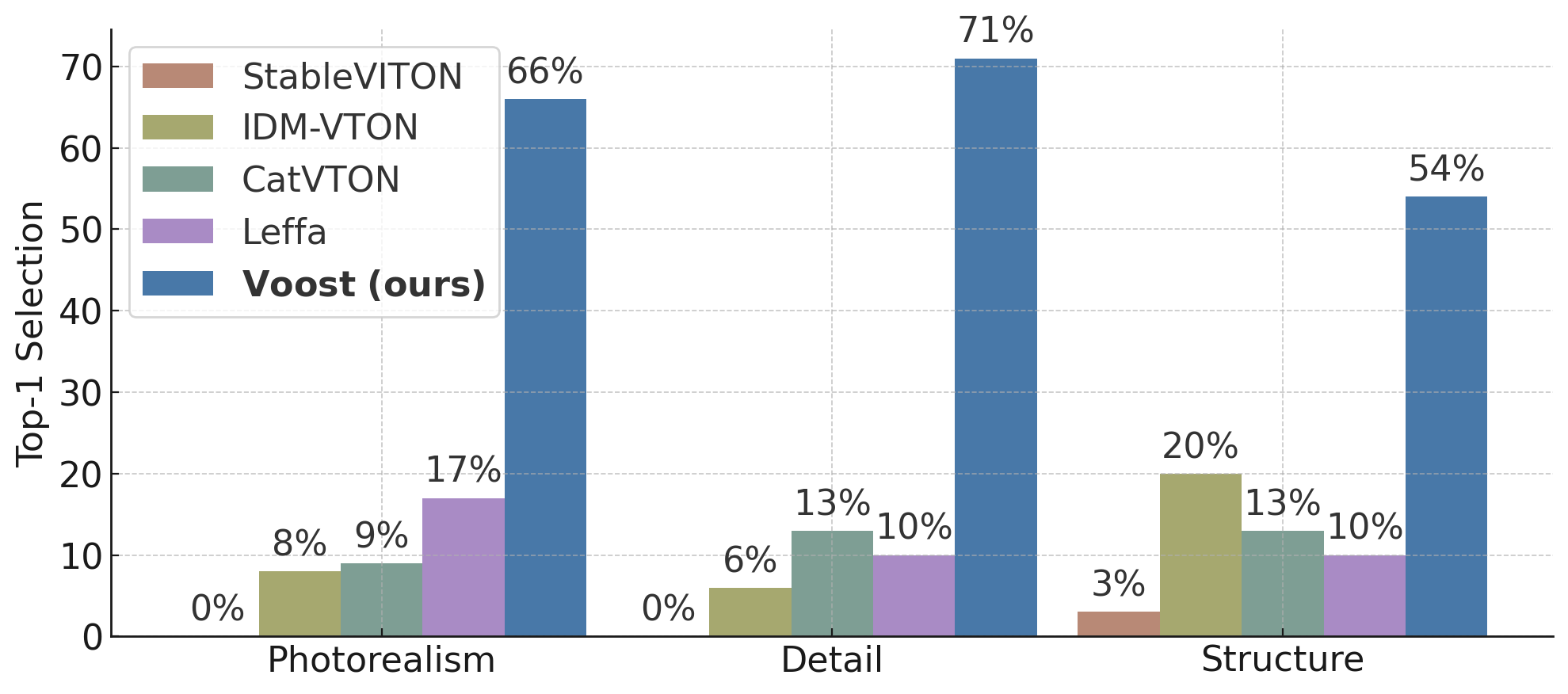}
    \caption{User study results. Comparing our method (\emph{Voost}) with other baselines on three criteria: photorealism, garment detail, and garment structure. Across all categories, \emph{Voost} received the highest top-1 selection rates, indicating clear user preference for its visual realism and garment fidelity.}
    \label{fig:exp_user_study}
\end{figure}

\paragraph{Effect of inference-time refinement.}
We evaluate the two inference-time strategies introduced in Sec.~\ref{subsec: inference}: temperature scaling and self-correction.
As shown in Table.~\ref{tab:quantitative_tryon}, temperature scaling provides consistent improvements on benchmarks like VITON-HD and DressCode, which feature relatively uniform image compositions and garment placements.
Its effectiveness becomes even more evident in challenging real-world scenarios. As illustrated in Fig.~\ref{fig:inference_temp_scaling}, it helps preserve garment fidelity when the masked region is small or spatially imbalanced.
The self-correction mechanism, while not included in the main quantitative benchmarks due to its selective and user-invoked nature, offers a practical enhancement in difficult cases.
As shown in Fig.~\ref{fig:inference_self_correction}, it can successfully recover garment structure and detail when initial generations fall short, improving perceptual realism without additional model retraining.

\begin{figure}[!t]
    \centering
    \includegraphics[width=\linewidth]{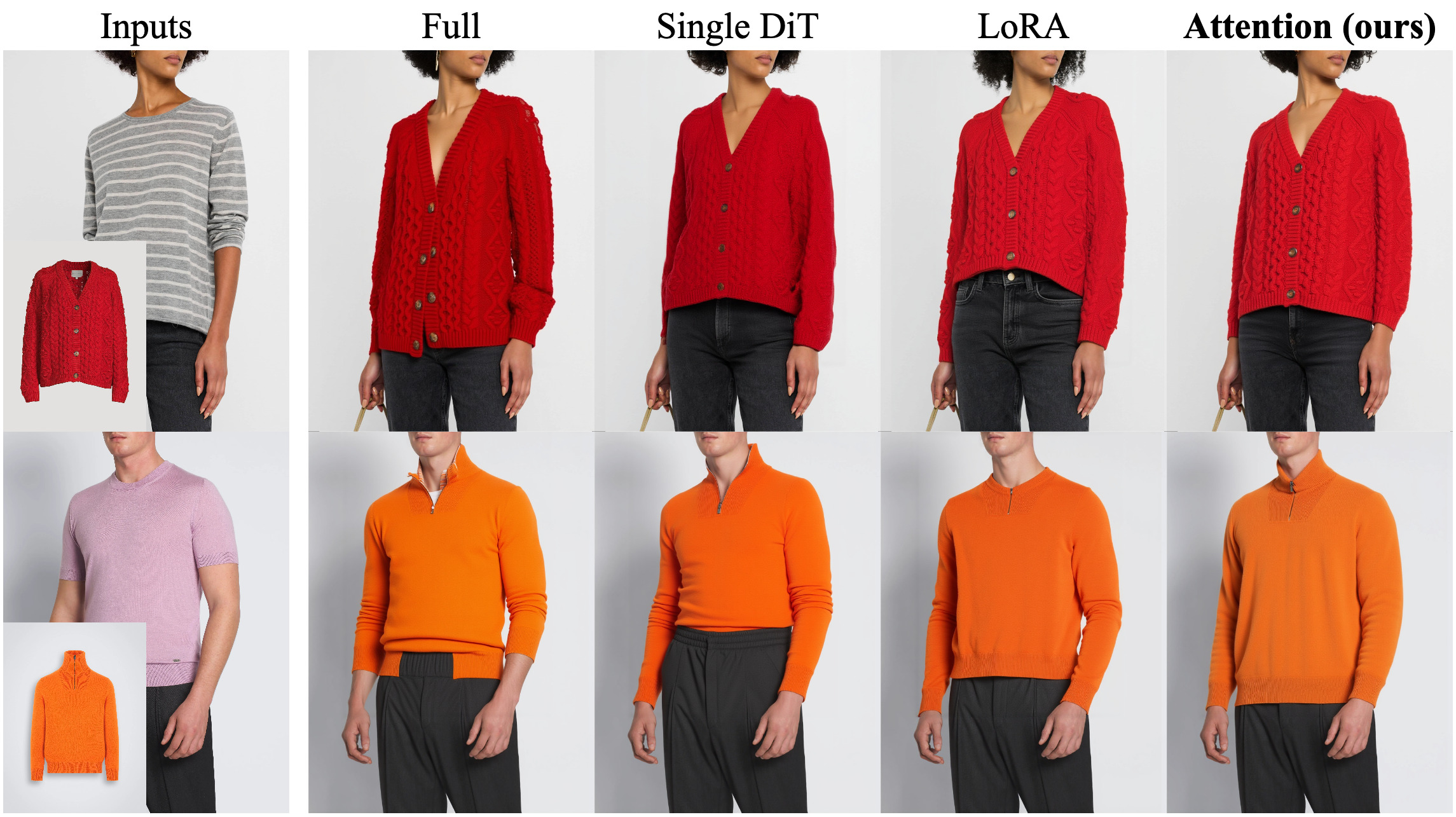}
    \caption{Qualitative comparison of training strategies.}
    \label{fig:exp_ablation_qual}
\end{figure}

\paragraph{Effect of trainable parameters.}
We compare training strategies with different subsets of trainable weights. As shown in Table~\ref{tab:train_strategy_comparison}, our attention-only tuning achieves the best overall performance, outperforming full fine-tuning, single-block training, and LoRA~\cite{hu2022lora}. It effectively captures garment–person correspondence while significantly reducing training cost.
Fig.~\ref{fig:exp_ablation_qual} further highlights the qualitative advantage: attention-only tuning yields sharper, more coherent outputs with fewer artifacts.
The \textit{supplementary material} provides a detailed analysis of this design choice, highlighting sensitivity studies that justify our focus on attention layers.

\section{Conclusion}
\label{sec:conclusion}

\begin{table}[t]
    \centering
    \scriptsize 
    \begin{tabular}{@{}l c cccc@{}}
        \toprule
        \textbf{Training strategy} & 
        \textbf{\# params.}
        & SSIM↑ & LPIPS↓ & FID↓ & KID↓ \\
        \midrule
        Full & 11.9B & 0.875 & 0.081 & 6.351 & 0.886 \\
        Single DiT Blocks & 5.38B & 0.872 & 0.078 & 5.975 & 0.634 \\
        \rowcolor{gray!20}
        Attention-only (\textbf{ours}) & \textbf{2.69B} & \textbf{0.899} & \textbf{0.056} & \textbf{5.269} & \textbf{0.404} \\
        LoRA (r=128) & 359M & 0.843 & 0.108 & 6.668 & 0.906 \\
        \bottomrule
    \end{tabular}
    \caption{Quantitative comparison of training strategies with different updated layers on VITON-HD~\cite{choi2021viton}.}
    \label{tab:train_strategy_comparison}

\end{table}

In this work, we proposed \emph{Voost}, a unified and scalable framework that jointly models virtual try-on and try-off within a single diffusion transformer. By formulating the two tasks as bidirectional counterparts, \emph{Voost} enables mutual supervision without relying on task-specific architectures, auxiliary losses, or additional labels.
To further improve robustness and consistency, we introduced two inference-time techniques: attention temperature scaling and self-corrective sampling. Comprehensive experiments demonstrate that \emph{Voost} consistently surpasses strong baselines on both try-on and try-off benchmarks, achieving state-of-the-art performance in alignment, visual fidelity, and generalization.
These results highlight the effectiveness of unified diffusion modeling for fashion synthesis and suggest a promising direction for integrated human–garment understanding.

\paragraph{Limitations and future work.}
While our model is capable of producing photorealistic results by jointly learning try-on and try-off through bidirectional attention, precise control over the garment's fit remains somewhat ambiguous due to the lack of explicit structural or sizing information.

In future work, we plan to incorporate additional cues such as body measurements or garment metadata to improve controllability. 
More broadly, the strong image-level foundation of \emph{Voost} makes it well-suited for downstream extensions such as video-based~\cite{karras2024fashion,he2024wildvidfit} or 3D-based~\cite{cao2024gs,he2025vton,zhang2025robust} synthesis, where consistent and faithful garment–person interaction remains essential yet challenging.


{
    \small
    \bibliographystyle{ieeenat_fullname}
    \bibliography{custom}
}
\clearpage
\clearpage
\maketitlesupplementary
\section{Implementation Details}

\subsection{Dataset composition}

We train our model on a combination of VITON-HD~\cite{choi2021viton}, DressCode~\cite{morelli2022dress}, and an additional curated set of real-world images. The latter consists of (1) high-resolution photographs captured in-house under controlled conditions and (2) in-the-wild fashion images from online sources. All external data were manually filtered to ensure high garment visibility, realistic lighting, and diverse poses.

Our dataset maintains a reasonably diverse distribution across clothing types and subject compositions. Among all garment inputs, 52.1\% are tops, 24.7\% are bottoms, and 23.2\% are full-body outfits. In terms of person image composition, 61.4\% depict women and 38.6\% depict men. Based on framing, 45.3\% are upper-body shots, 31.8\% full-body, and 22.9\% lower-body.

Unlike previous works that operate on fixed-size inputs, our model supports variable aspect ratios during both training and inference (see Section~3.2). Accordingly, the dataset includes a diverse range of aspect ratios. The most common format is 3:4 (width:height), followed by 1:1, 2:3, 1:2, and 1:3, with a smaller portion falling outside these standard categories. This diversity enables our model to generalize well to in-the-wild scenarios with arbitrary image dimensions.

\subsection{Mask augmentation}

To prevent overfitting and improve generalization, we apply targeted augmentations to the agnostic masks, using distinct strategies for try-on and try-off tasks.

For \textbf{try-on}, we ensure that all garment regions in the person image are fully masked, while explicitly preserving facial and hand regions. To obtain these masks, we leverage state-of-the-art human parsing and pose estimation models~\cite{kirillov2023segany,li2019selfcorrectionhumanparsing,guler2018densepose,khirodkar2024sapiens}. During training, we generate diverse masks of varying shapes and sizes that perfectly occlude the clothing area but retain key human features. This encourages the model to rely entirely on the conditioning garment image rather than residual cues in the person input, significantly improving silhouette realism and garment fit, especially for longer garments.

For \textbf{try-off}, we use a dichotomous segmentation model~\cite{meyer2025benusingconfidenceguidedmatting} to isolate the garment region, and apply masks of varying sizes—from tight garment-specific masks to nearly full-image masks. This approach forces the model to infer plausible underlying appearances across a range of masking levels, enhancing robustness under challenging conditions.

\subsection{Optimization}

We use the AdamW~\cite{loshchilov2018decoupled} optimizer with a learning rate of $1 \times 10^{-5}$ and a weight decay of 0.001. The model is trained with a batch size of 128 using the DeepSpeed ZeRO-2~\cite{rasley2020deepspeed} optimizer state partitioning strategy for approximately 48 single-H100 GPU days.

\subsection{Inference}

At test time, we employ a flow-matching-based Euler scheduler introduced in Stable Diffusion 3~\cite{esser2024scalingrectifiedflowtransformers}, using exactly 28 sampling steps. All qualitative results shown in the supplementary figures were generated using this consistent inference setup.

\subsection{Inference Technique Details}

\paragraph{Temperature Scaling.}
We use $\alpha = 1.0$ and $\beta = 0.43$ in our temperature scaling formula. The value of $\beta$ was empirically chosen based on garment-to-mask ratios aggregated over a large number of training samples, capturing typical spatial layouts seen during training.

\paragraph{Self-Correction.}
For self-corrective sampling, we apply updates at two timesteps, $t=5$ and $t=17$, based on a total denoising schedule of $T=28$. These correspond to early and mid-to-late stages of the denoising process, chosen to target complementary aspects: $t=5$ primarily influences coarse shape, while $t=17$ contributes to fine detail refinement.
At the shape correction step ($t=5$), we use the same mask as the standard try-off setting, $\mathbf{M} = [\mathbf{1} \mid \mathbf{0}]$, encouraging the model to recover overall garment length and structure from the try-on result. For detail correction ($t=17$), we instead use a garment-specific mask obtained via segmentation models~\cite{meyer2025benusingconfidenceguidedmatting,kirillov2023segany,khirodkar2024sapiens} of the input garment image. This focuses the gradient signal on texture and detail consistency, while avoiding penalization for plausible geometric differences.
We set the number of refinement iterations to $R=5$ for each correction step.

\begin{figure*}
    \centering
    \includegraphics[width=\linewidth]{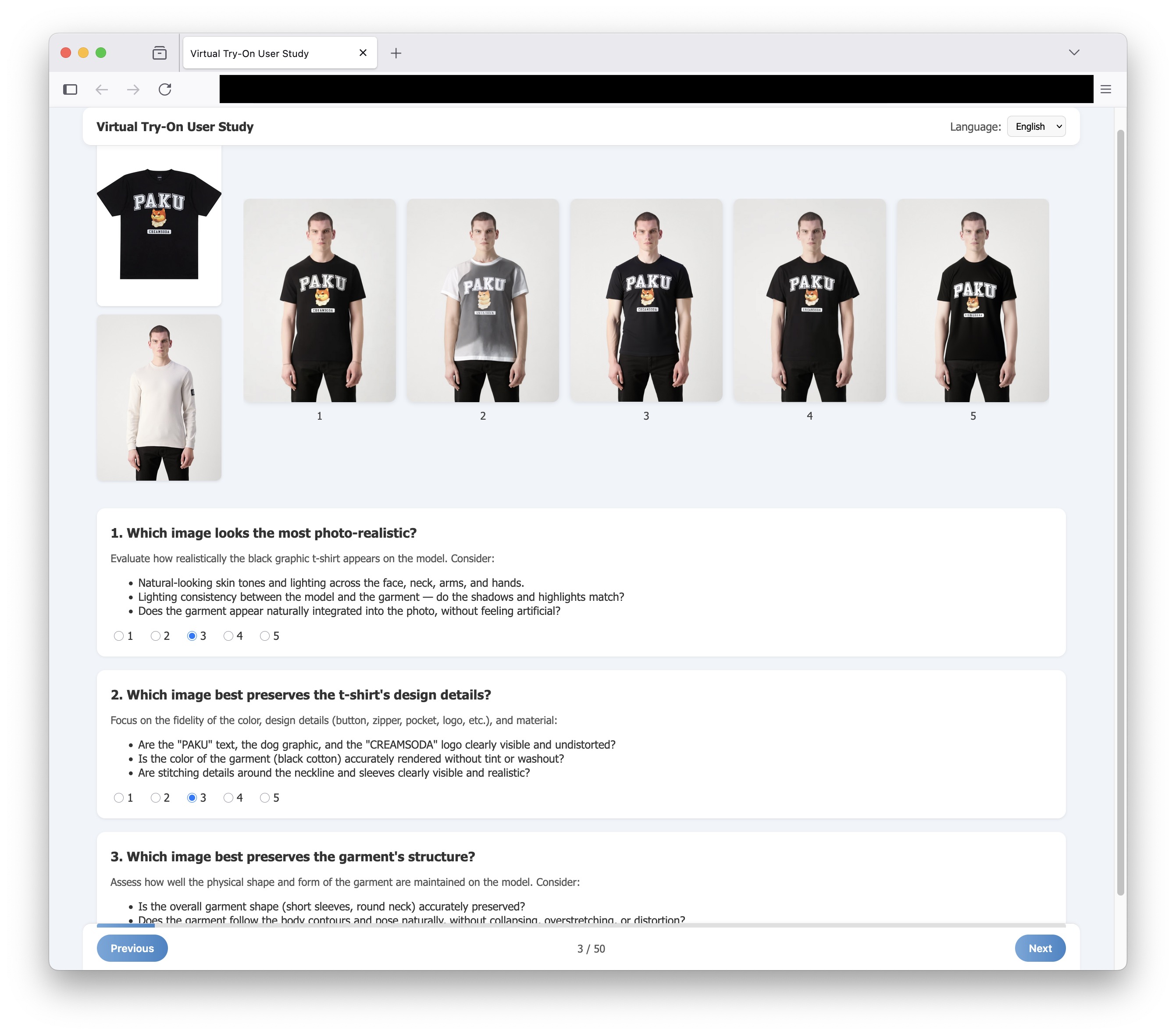}
    \caption{
    User study interface for evaluating photorealism, design detail, and garment structure across five try-on results.
    }
    \label{suppl:user_study}
\end{figure*}


\begin{figure}[t]
    \centering
    \includegraphics[width=0.9\linewidth]{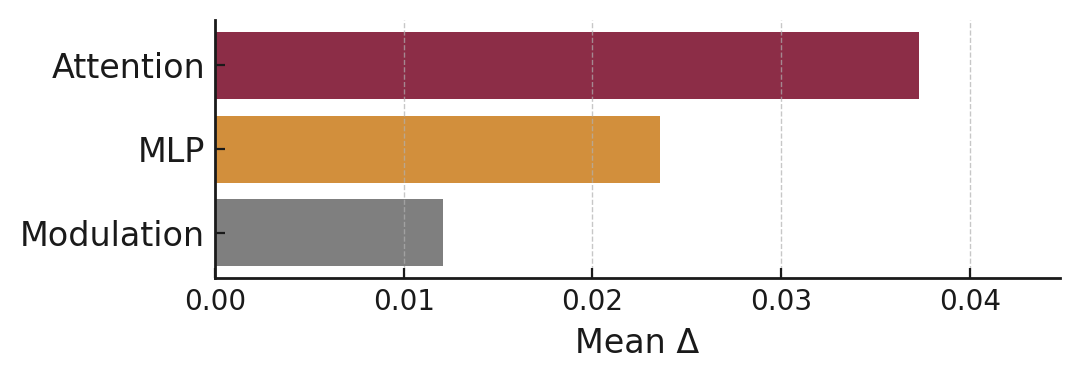}
    \caption{Layer-wise parameter update magnitudes during full parameter tuning.}
    \label{fig:param_change_analysis}
\end{figure}
\section{Rationale for Attention-Only Training}

To adapt a pretrained DiT for virtual try-on and try-off under our concatenation-based setup, we fine-tune only the attention modules while freezing all other parameters—a choice grounded in both empirical evidence and task-specific insight.

Following~\cite{li2020fewshotimagegenerationelastic, kumari2023multiconceptcustomizationtexttoimagediffusion}, we analyze parameter sensitivity by fully fine-tuning the model on concatenated inputs $[\mathbf{X}_{\mathrm{g}} \mid \mathbf{X}_{\mathrm{p}}]$ and computing the normalized weight update for each layer as
\[
\Delta_l = \lVert \boldsymbol{\theta}'_l - \boldsymbol{\theta}_l \rVert / \lVert \boldsymbol{\theta}_l \rVert,
\]
where $\boldsymbol{\theta}_l$ and $\boldsymbol{\theta}'_l$ denote the pretrained and updated weights, respectively, and $\lVert \cdot \rVert$ is a general vector norm. As shown in Fig.~\ref{fig:param_change_analysis}, attention layers exhibit the most significant changes, underscoring their key role in garment-to-person alignment.

In contrast, full-model fine-tuning often leads to overfitting, introducing spurious accessories or artifacts unrelated to the garment. This suggests that unconstrained updates may distort the conditioning signal and reduce generation fidelity.

By limiting updates to attention layers, we preserve the pretrained diffusion prior while enabling controlled, localized adaptation—achieving a favorable trade-off between flexibility and stability.

\section{Details on Human Evaluation}

This section provides details on the user study protocol described in Sec. 4.4. Participants were shown a reference garment and person image, along with five generated try-on results from our model and four from other state-of-the-art baselines.

Each result was evaluated using three questions:
(1) Which image looks the most \textbf{photo-realistic}?
(2) Which image best preserves the garment's \textbf{design details}?—Focus on the fidelity of the color, design details (button, zipper, pocket, logo, etc.), and material
(3) Which image best preserves the garment's \textbf{structure}?—Assess how well the physical shape and form of the garment are maintained on the model.

We evaluated 50 different garment-person pairs, yielding 150 unique queries. Each query was answered by 30 independent users, resulting in a total of 4,500 responses. To ensure fairness, we provided specific assessment guidelines for every 150 query—for example, \textit{whether text or graphics (e.g., “PAKU”, a dog graphic, or “CREAMSODA” logo) appeared clearly and without distortion.}

A screenshot of the evaluation interface is shown in Fig.~\ref{suppl:user_study}, including instructions and visual examples.

\section{Failure Cases}
\begin{figure}[!t]
    \centering
    \includegraphics[width=\linewidth]{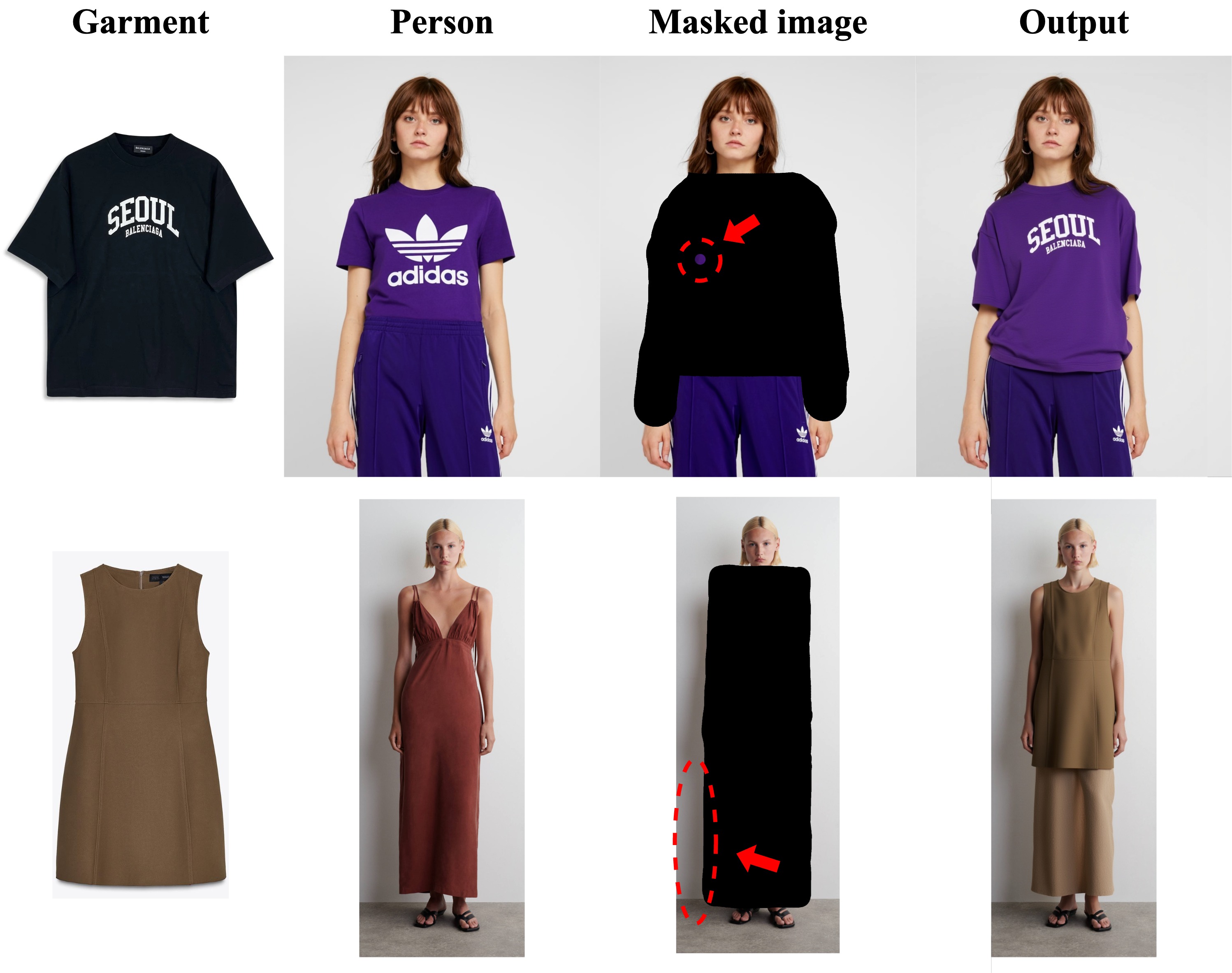}
   \caption{
Failure cases of our method. 
(\textbf{Top}) When the input mask, either user-provided or automatically generated, does not fully cover the original garment, the model may use the exposed regions as inpainting priors, resulting in undesired continuation of the original clothing. 
(\textbf{Bottom}) Even with a correct mask, residual cues such as shadows from the original garment can bias the generation process, leading to incomplete or distorted reconstruction of the target garment.
}
    \label{fig:failure_case}
\end{figure}

While our method demonstrates superior and robust performance across various conditions, there are certain failure cases worth noting (Fig.~\ref{fig:failure_case}). 

One common issue occurs when the input mask provided by the user or generated automatically does not fully cover the original garment. 
In such cases, the model leverages the exposed regions as strong inpainting priors, leading to undesired continuation of the original garment instead of fully replacing it with the target garment. 

We also observe cases where, even with a correctly defined mask, subtle cues such as shadows or shading patterns from the original clothing bias the generation process, resulting in incomplete or distorted reconstruction of the intended garment.

These cases highlight the sensitivity of diffusion-based inpainting to residual visual cues and motivate future work on more robust garment masking and conditioning strategies.

\section{Extra Qualitative Results}

\begin{figure*}
    \centering
    \includegraphics[width=\linewidth]{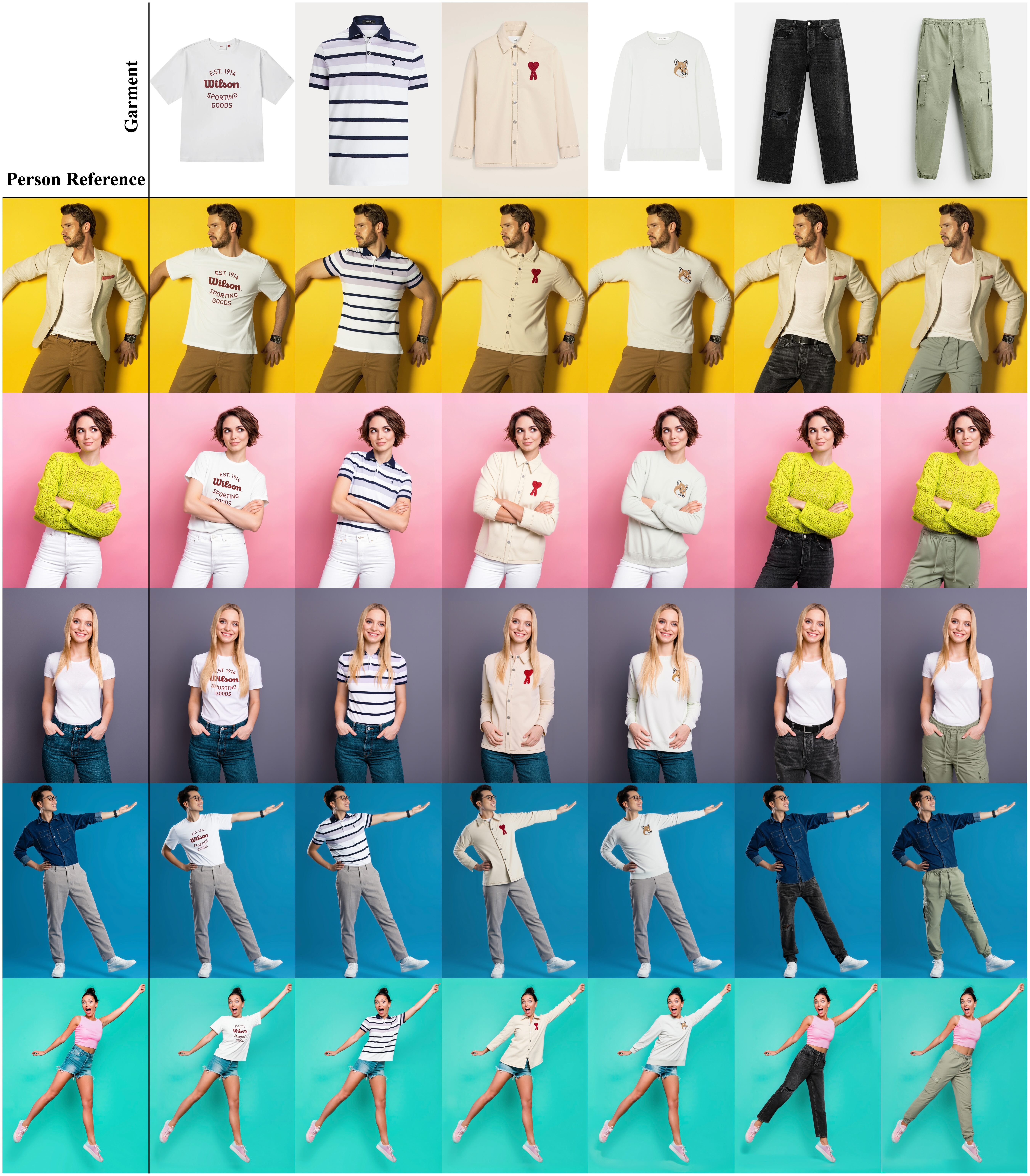}
    \caption{Qualitative comparison of try-on results in the wild, showcasing various clothing styles, poses, and viewpoints.}
    \label{suppl:qual_cross}
\end{figure*}

\begin{figure*}
    \centering
    \includegraphics[width=0.92\linewidth]{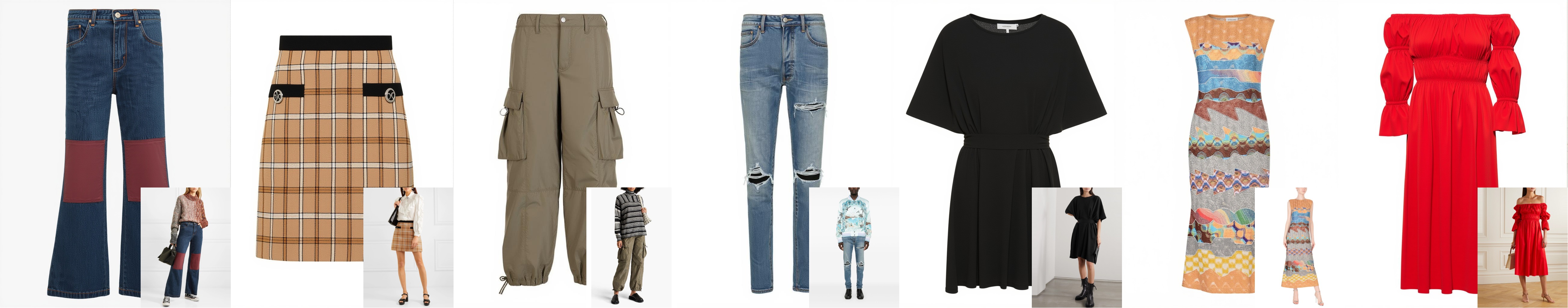}
    \caption{Qualitative results of the try-off task on the in-the-wild images, demonstrating performance across various clothing types including lower and full-body clothing.}
    \label{suppl:qual_tryoff_ours}
\end{figure*}

We further showcase the versatility of our model in Fig.~\ref{suppl:qual_cross}, ~\ref{suppl:qual_tryoff_ours} which contains additional try-on and try-off results across a wide range of garments and human appearances. The samples include diverse garment types applied to people with varying poses, viewpoints, and body shapes. These results highlight the robustness of our unified model in producing perceptually convincing outputs under challenging real-world conditions.

\section{More Qualitative Comparisons}

We present additional qualitative comparisons with state-of-the-art baselines on VITON-HD~\cite{choi2021viton}, DressCode~\cite{morelli2022dress}, and a set of in-the-wild examples gathered from online sources.

\begin{figure*}[!t]
    \centering
    \includegraphics[width=\linewidth]{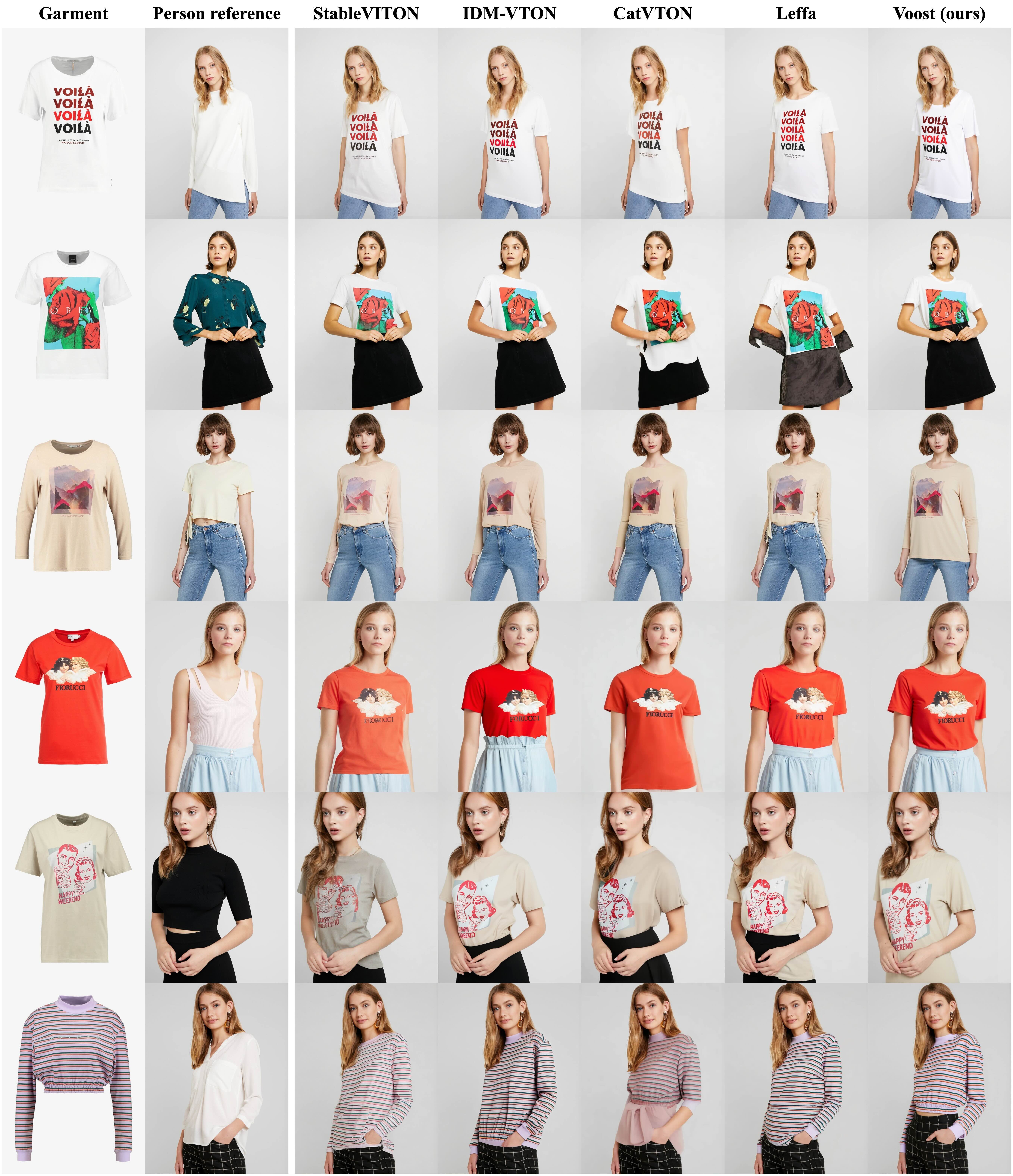}
    \caption{Qualitative comparison of virtual try-on results on the VITON-HD dataset.}
    \label{suppl:comp_viton}
\end{figure*}

\begin{figure*}
    \centering
    \includegraphics[width=\linewidth]{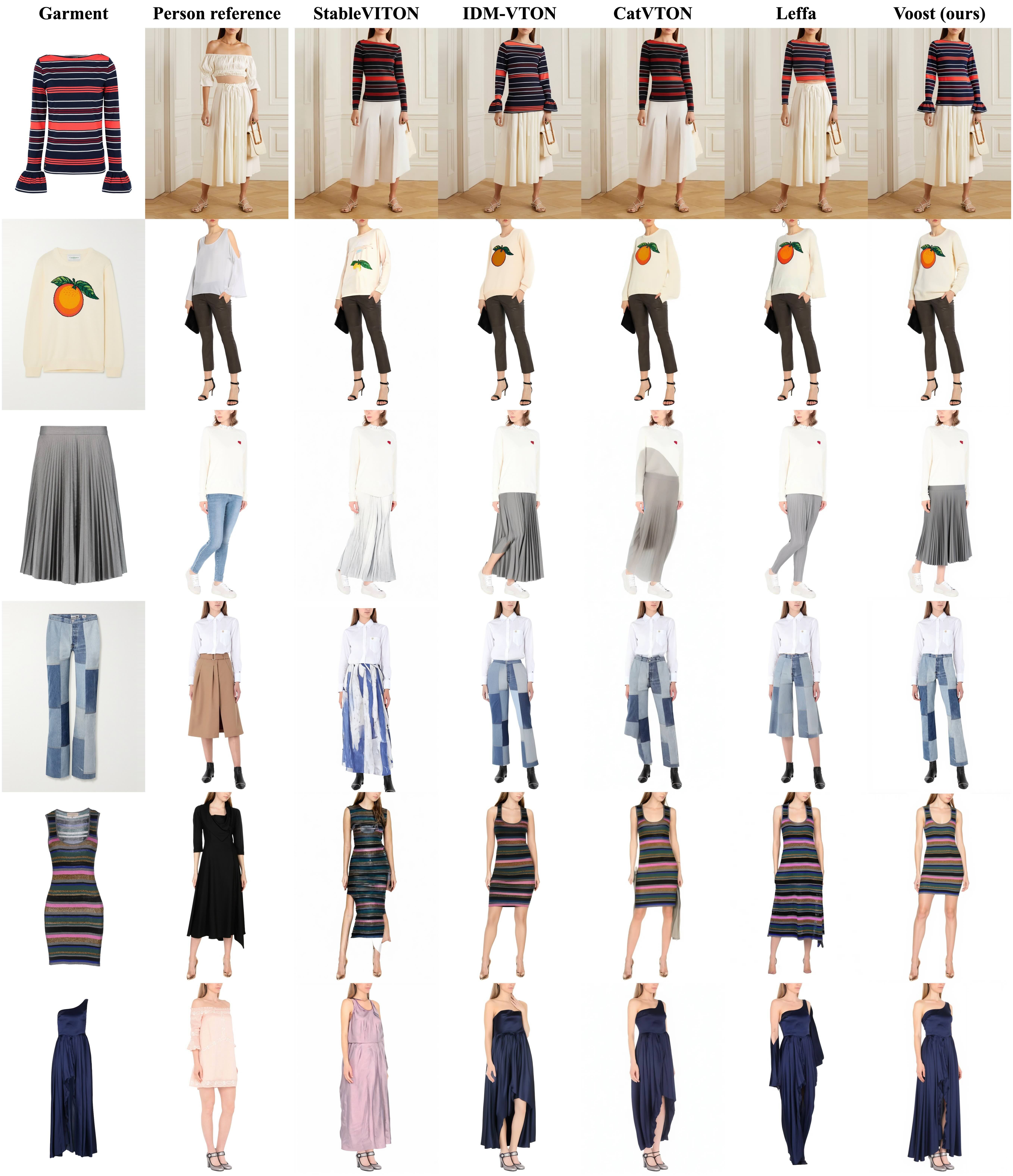}
    \caption{Qualitative comparison of virtual try-on results on the DressCode dataset.}
    \label{suppl:comp_dresscode}
\end{figure*}

\begin{figure*}
    \centering
    \includegraphics[width=\linewidth]{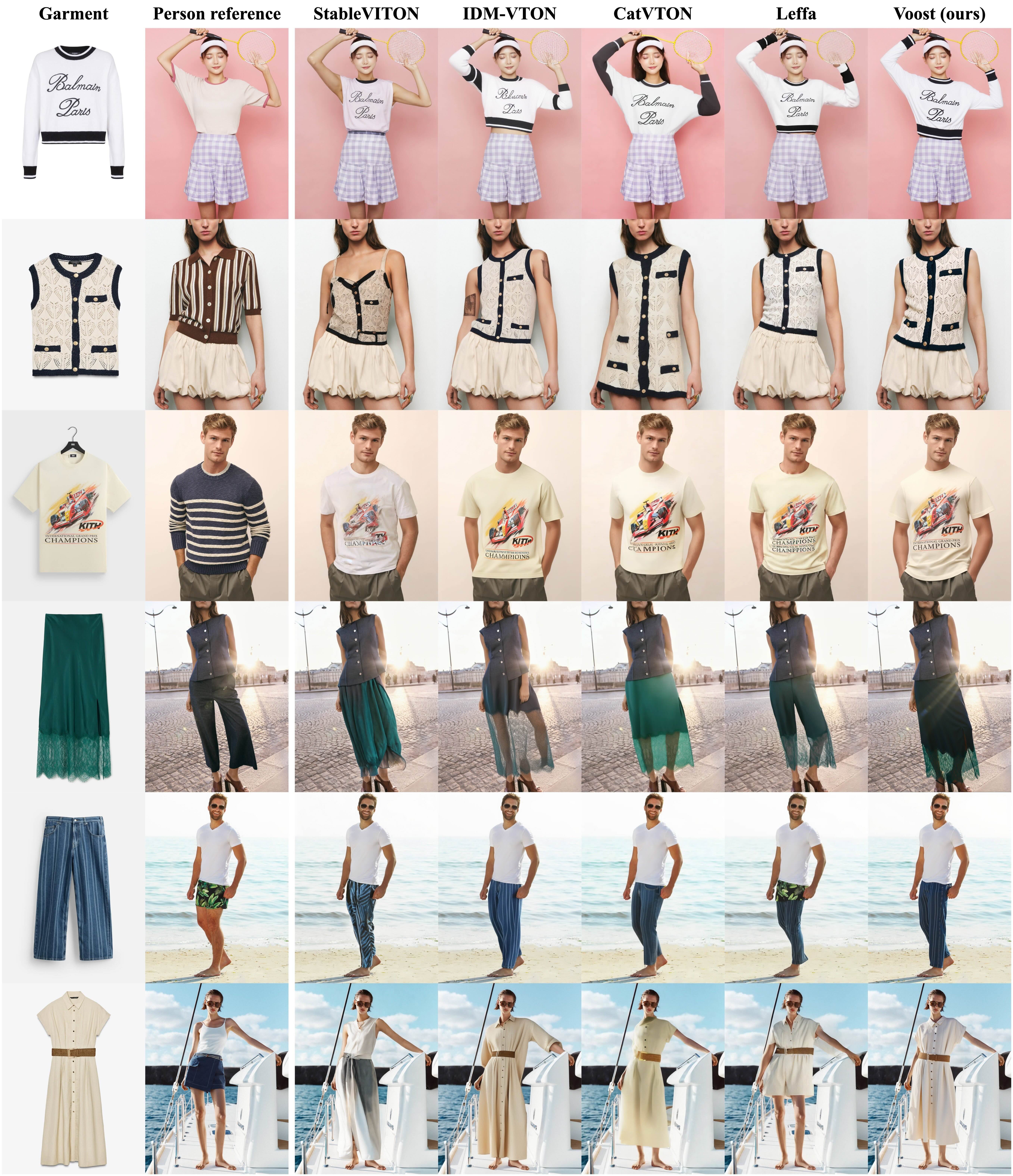}
    \caption{Qualitative comparison of virtual try-on results on the in-the-wild images.}
    \label{suppl:comp_inthewild}
\end{figure*}

Fig.~\ref{suppl:comp_viton} shows try-on results on VITON-HD, highlighting our model’s ability to maintain structural consistency and garment fidelity. Fig.~\ref{suppl:comp_dresscode} presents results on DressCode, demonstrating improved alignment and design preservation across various clothing types. To further evaluate generalization, Fig.~\ref{suppl:comp_inthewild} compares in-the-wild images, where our model consistently generates realistic and artifact-free outputs under diverse poses, lighting, and backgrounds.






\begin{figure*}
    \centering
    \includegraphics[width=0.95\linewidth]{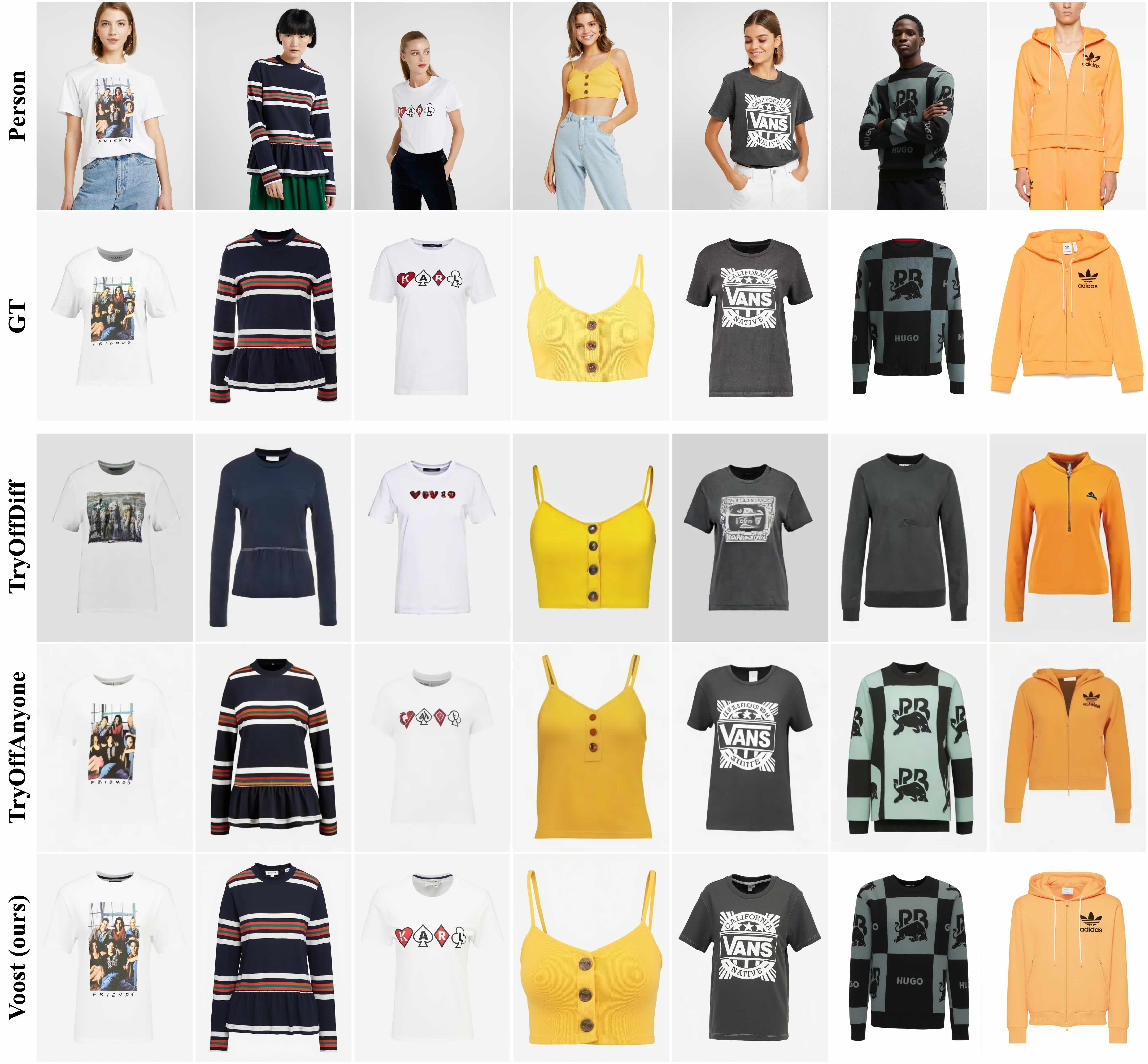}
    \caption{Qualitative comparison of try-off results on the VITON-HD and in-the-wild images. For comparison with other methods, the evaluation is conducted using upper clothing only.}
    \label{suppl:qual_tryoff}
\end{figure*} 

\end{document}